\documentclass [11pt]{article}
\usepackage{amsmath,amsthm,amsfonts,amscd,eucal,latexsym,amssymb,mathrsfs}
\usepackage{epsfig}  
\def\cB{{\ca B}}

\def\cD{{\ca D}}
\def\cE{{\ca E}}
\def\cF{{\ca F}}

\def\cH{{\ca H}}

\def\cO{{\ca O}}

\def\cS{{\ca S}}
\def\cT{{\ca T}}

\def\bC{{\mathbb C}}           

\def\bM{{\mathbb M}} 
 
\def\bR{{\mathbb R}}

\def\bS{{\mathbb S}}

\newsymbol\rest 1316         
 
\def\mA{\mathscr{A}} 
\def\mC{\mathscr{C}}

\def\mF{\mathscr{F}}  
\def\mI{\mathscr{I}}

\def\mB{\mathscr{B}}

\def\mT{\mathscr{T}}

\def\beq{\begin{eqnarray}}
\def\eeq{\end{eqnarray}}
\def\pa{\partial}
\def\at{\left(}               
\def\aq{\left[}               
\def\ag{\left\{}              

\def\ct{\right)}              
\def\cq{\right]}              
\def\cg{\right\}}             
\newcommand{\ca}[1]{{\cal #1}}         

\def\ga{\gamma}
\def\de{\delta}

\def\ka{\kappa}
\def\la{\lambda}

\def\si{\sigma}
\def\om{\omega}
\def\bom{{\boldsymbol{\omega}}}

\def\De{\Delta}

\def\Si{\Sigma}
\def\Om{\Omega}

\def\scrim{{\Im^-}}

\def\DX{\de X}

\newcommand{\nref}[1]{(\ref{#1})}

\newcommand{\se}[1]{\section{#1}}

\def\vsp{\vspace{0.2cm}}

\def\sse #1 {\vsp\ifhmode{\par}\fi\refstepcounter{subsection}
  \noindent {\bf\thesubsection}. {\em #1}.\quad
  \addcontentsline{toc}{subsection}{\protect\numberline{\thesubsection} #1}%
  }

\def\ssb #1 {\vsp\ifhmode{\par}\fi\refstepcounter{subsection}
  \noindent {\bf\thesubsection.} {\bf #1.}\quad
  \addcontentsline{toc}{subsection}{\protect\numberline{\thesubsection} #1}%
  }

\def\ssa #1 {\ifhmode{\par}\fi\refstepcounter{subsection}
  \noindent {\bf\thesubsection.} {\bf #1.}\quad
  \addcontentsline{toc}{subsection}{\protect\numberline{\thesubsection} #1}%
  }

\def\remark {\vsp\ifhmode{\par}\fi\noindent\noindent {\bf Remark:} 
}

\newtheorem{teorema}{Theorem}[section]
\newtheorem{proposizione}{Proposition}[section]

\newtheorem{definizione}{Definition}[section]

\begin{document}

\hfill{\sl Desy Preprint 10-002 - ZMP-HH/10-1 - January 2010} 
\par 
\bigskip 
\par 
\rm 
 
 
\par 
\bigskip 
\LARGE 
\noindent 
{\bf On the initial conditions and solutions of the semiclassical Einstein equations in a cosmological scenario.}
\bigskip 
\par 
\rm 
\normalsize 
 

\large
\noindent {\bf Nicola Pinamonti} 

\par
\small
\noindent 
II. Institut f\"ur Theoretische Physik, Universit\"at Hamburg,
Luruper Chaussee 149, 
D-22761 Hamburg, Germany.\smallskip

\noindent 
Dipartimento di Matematica, Universit\`a di Roma ``Tor Vergata'', Via della Ricerca Scientifica, I-00133 Roma, Italy.\smallskip

\noindent  nicola.pinamonti@desy.de\\ 
 \normalsize

\par 
 

\rm\normalsize 
 
 
\par 
\bigskip 

\noindent 
\small 
{\bf Abstract}.
In this paper we shall discuss the backreaction of a massive quantum scalar field on the curvature, the latter treated as a classical field. 
Furthermore, 
we shall deal with this problem in the realm of cosmological spacetimes by analyzing the Einstein equations in a semiclassical fashion.
More precisely, we shall show that, at least on small intervals of time, solutions for this interacting system exist.  
This result will be achieved providing an iteration scheme and showing that the series, obtained starting from the massless solution, converges in the appropriate Banach space.
The quantum states with good ultraviolet behavior (Hadamard property), used in order to obtain the backreaction, 
will be completely determined by their form on 
the initial surface if chosen to be lightlike. Furthermore, on small intervals of time, they do not influence the behavior of the exact solution.
On large intervals of time the situation is more complicated but, if the spacetime is expanding, we shall show that the end-point of the evolution does not depend strongly on the quantum state, because, in this limit, the expectation values of the matter fields responsible for the backreaction do not depend on the particular homogeneous Hadamard state at all.
Finally, we shall comment on the interpretation of the semiclassical Einstein equations  for this kind of problems.
Although the fluctuations of the expectation values of pointlike fields diverge, if the spacetime and the quantum state have a large spatial symmetry and if we consider the smeared fields on regions of large spatial volume, they tend to vanish. Assuming this point of view  the semiclassical Einstein equations become more reliable.

\normalsize
\bigskip 

\tableofcontents

\se{Introduction} 

Despite the presence of many attempts, no widely accepted  description of the quantization procedure necessary to treat quantum gravity is, up to now, available; 
however, nowadays we know how such a theory should look like, at least in some regimes. 
For example, treating fields as quantum object over a fixed curved classical background, many interesting effects have been discovered, like for example Hawking radiation in the case of black hole physics \cite{Hawking, FH} or particle creation in the contest of cosmological spacetime \cite{Parker1,Parker}.
Another interesting regime, where some progress could be made towards an eventual quantum gravity, is the analysis of the backreaction of matter quantum fields on curvature. 
In the present paper we would like to analyze some aspects of this problem in the context of cosmological spactimes.

In this respect the recent developments in the theory of quantum fields on curved spacetimes provided by the algebraic approach to quantization \cite{Haag}, have been really fruitful.
Here we think for example at the work \cite{BFV} where it has been shown, that the quantization is nothing but a functor that associates to every spacetime its corresponding algebra of fields/observables.
It is thus possible to quantize simultaneously on every spacetime and, as a byproduct, the renormalization ambiguities, present in the definition of  pointwise product of fields, acquire a clearer interpretation \cite{HW01}.
In order to use similar ideas, 
a good control on the form of the quantum states has to be available   and, in this respect, an important contribution is the one of Radzikowksi \cite{Radzikowski} (see also \cite{BFK}).
In that paper he has shown that, considering a free quantum theory and 
employing techniques proper of the H\"ormander microlocal analysis \cite{Hormander}, 
it is sufficient to prescribe the form of the wave front set
of the two-point function of a state in order to  fix the form of its singularity (Hadamard form).
Using similar ideas and similar methods  it is also possible to show that many of the features of quantum fields on flat spacetime can be translated in a conceptual clear way to the curved case.
More precisely, the Wick polynomial, their time ordered products and the stress tensor have been addressed in
\cite{HW01,HW02, HW05,Mo03}, whereas the study of the perturbative construction of interacting field theories can be found in \cite{BF00, HW03, BF, BDF}.

Coming back to the analysis of the backreaction of quantum matter on classical curvature, 
a milestone is provided by the seminal paper of Wald
\cite{Wald77}.
In that work, he has given a set of five axioms that a renormalization scheme should satisfy in order to produce a stress tensor whose expectation values can be safely used in the semiclassical Einstein equations 
	\beq\label{semiclassical1}
	G_{\mu\nu}=8\pi \langle T_{\mu\nu}\rangle
	\eeq
in such a way that a physically reasonable backreaction emerges.
Nowadays, the first four requirements on the expectation values of the stress tensor (consistency,
covariance, conservation of $\langle T_{\mu\nu}\rangle$ and standard flat spacetime limit) 
 are automatically obtained in every locally covariant theory constructed as in \cite{BFV} (see also the nice lectures \cite{BF}).
However, the fifth one, which requires that the expectation values of the stress tensor depend on the derivatives of the coefficients of the metric up to the second order is more problematic. Nevertheless, even if it cannot be satisfied in general, it is possible to fix part of the renormalization freedom in order to have that at least the expectation values of the trace $\langle T\rangle:=\langle {T_\mu}^\mu \rangle$ 
meet this requiremeant \cite{Wald78}.

Making use of these methods, in the case of cosmological spacetimes, the backreaction of a massless conformally coupled scalar field has been studied by Wald in \cite{Wald78a} and in this case an exact solution can be found.
Later on, another important contribution in this direction is the one due to Starobinsky
\cite{Starobinsky} (see also \cite{Vile85}), where analyzing the trace of the stress tensor, and its anomalous term \cite{Wald78},  the backreaction on curvature was computed.
Following similar ideas the analysis of the backreaction of massive scalar fields has been analyzed in
\cite{Shapiro} and more recently in \cite{DFP}. 
Unfortunately while in the case of massless conformally coupled fields the backreaction can be analytically computed, essentially because the quantum state does not influence the dynamics of the spactime, when the fields have a mass, the situation is more complicated and the 
details of the quantum state have to be taken into account.
A nice study of the semiclassical problem, also in terms of numerical integration, can be found in the series of works of Anderson \cite{Anderson3, Anderson4}.
The analysis of the problem of definition of the expectation values of the stress tensor in relation with the Casimir effect is presented in  
\cite{GNW}.
The semiclassical Einstein equations have also been employed by Flanagan and Wald in \cite{FW} in order to study the validity of the averaged null energy condition on semiclassical solutions.
Other interesting works at the interface of classical gravity and quantum matter are \cite{GottloberMuller, RV, ParkerSimon, ParkerRaval, AndersonEaker, Shapiro1}.
Recently the dynamical system associated with the semiclassical backreaction has been studied in \cite{EG}, where similar equations, as the one studied in the present paper, have been derived also for more general kinds of matter fields (with different curvature coupling).

In this paper we shall follow the discussion presented in \cite{DFP} where some approximated solutions of the semiclassical Einstein equations have been displayed. In the latter work the matter has been described by a free scalar field whose coupling constant to the curvature is equal to $1/6$ and with a non vanishing mass $m$.
Notice that, despite the simplicity of the treated model, the solutions presented in that paper have very nice physical properties. On the one hand they show, at late times, a phase of exponential expansion 
(de Sitter), whose acceleration corresponds to a renormalization freedom of such a model, while on the other hand, at the beginning of the model a phase typical of the power law inflationary scenario emerges.
In a certain sense that model provides an effective cosmological constant that varies in time and that could be used in order to model the present acceleration of our universe.

Here, we shall show that exact solutions of that model exist and 
that many of the previously discussed physical properties shown by the approximated solutions persist.
To this end, since we are here interested in the cosmological scenario we shall consider the category\footnote{The elements of $\cF$ are the oriented and time oriented flat FRW spacetimes and the morphisms can be chosen to be the orientation preserving conformal transformations} $\cF$ of spacetimes with a large spatial symmetry namely the flat Friedmann Robertson Walker (FRW).
Thanks to the result stated in \cite{BFV} and \cite{HW01}, the fields realizing the components of the stress tensor can be unambiguously found on every spacetime, hence also on every element of $\cF$. However,
in order to have a well defined system of equations governing the  backreaction, we must be able to 
unambiguously select a class of states (one for every element of $\cF$) with good ultraviolet behavior in a spacetime independent way.
Eventually, the validity of the semiclassical Einstein equations will be employed as a constraint on the class $\cF$.
The difficult part in this program is 
to give, for every element of $\cF$, a state which is not based on the particular spacetime.
%
Here we would like to give a prescription for obtaining those natural states.
Furthermore, in order to give meaning to observables like the stress tensor, the states must have good ultraviolet behavior, namely the singular part of their two-point function must be as close as possible to the singularity shown by the vacuum in Minkowksi spacetime.
Unfortunately this is not always possible and in many cases, a state, determined by some initial values on a spacelike Cauchy surface, does not satisfy the necessary regularity conditions.
From the physical point of view, a natural candidate for a quantum state would be the choice of a vacuum, but unfortunately such concept is not available 
in curved spacetime. 
In any case
we could ask for the validity of some physical properties in order to select a class of states that behaves like the vacuum in Minkowski spacetime, or at least that are close to it.
To this end, Parker \cite{Parker} introduced the adiabatic states, namely states that are close to the vacuum. Afterwards,  L\"uders and Roberts \cite{LuRo} made their definition precise employing the  construction of Parker to select certain initial conditions at finite time in a FRW spacetime.
Unfortunately, in general, the states obtained in that way do not satisfy the Hadamard condition, hence they do not have 
the same ultraviolet singularity as the vacuum in Minkowski spacetime, 
as shown by Junker and Schrohe 
 \cite{JuSc}; nevertheless 
 they satisfy a condition weaker than the microlocal spectral condition given in \cite{BFK}.
 Another possibility would be to consider states that minimize the smeared energy density.  Olbermann \cite{Olbermann} has constructed such states and has shown that they are of Hadamard form.
Unfortunately, since this procedure depends upon the smearing over a finite interval of time, 
it is not possible to get those states only out of the initial conditions.

In the present work we shall follow another procedure close to the one presented in \cite{DMP, Moretti06, Moretti} in the context of asymptotically flat spacetimes and already employed in cosmological scenarios in \cite{DMP2,DMP4}.
We shall see that, if the initial surface is chosen to be  lightlike (coinciding with the null past infinity in the conformally related spacetime), we can give a class of states (that do not rely on the particular spacetime) that can be used in order to solve the coupled semiclassical system.
Eventually, we shall comment on the late time behavior of the solutions showing that, if a de Sitter phase of exponentially accelerated expansion has emerged, the value of that acceleration does not depend on the particular homogeneous Hadamard state.
Finally we shall give a new interpretation of the semiclassical Einstein equations when considered as a non local equation. We shall actually show that, when smeared on large spatial volume regions, while the expectation value of the trace, $\langle T \rangle$, does not differ from the pointlike one, its fluctuations tend to vanish.

In the subsequent section we shall introduce the geometric setup and the basic facts about the quantization we shall use in the next, this section is used to fix the notation and to recollect some results used throughout the subsequent part of the work.
In the third section a discussion about the regularity and the Hadamard property shown by a class of states will be discussed, the states constructed in a spacetime independent way are presented in this section.
The fourth section contains the main theorems about the existence of solutions of the semiclassical Einstein equations in a cosmological scenario. In the fifth section we shall discuss the behavior of the semiclassical solutions at late times and we shall see that the variance of the trace vanish when smeared on large volume regions. Finally, some comments and remarks are collected in the sixth section.
The appendix contains a discussion about deformation quantization, the technical estimates and a technical proof of a proposition used in the main text.

\se{Geometry of the spacetime, classical and quantum matter}

\subsection{Geometry}
Since we are dealing with problems in cosmology we shall assume homogeneity and isotropy.  Hence, every element $(M,g)$ in the class of spacetimes  under investigation is such that the smooth manifold $M$ is of the form $M= I\times \Si$ where $I\subset \bR$ is an interval representing the {\bf cosmological time} and $\Si$ is a three dimensional hypersurface that can be closed open or flat.
The metric $g$, takes the well known Friedmann-Robertson-Walker form:
\beq\label{FRW}
g=-dt^2+a^2(t)\left[\frac{dr^2}{1-\ka r^2}+r^2d\bS^2(\theta,\varphi)\right],
\eeq
where $d\bS^2(\theta,\varphi)$ is the standard metric of the unit sphere in spherical coordinates and $\ka\in\{+1,0,-1\}$ distinguishes between the topologies of $\Si$, in particular between  
 open, flat or closed spacelike hypersurfaces respectively.
Furthermore, $a(t)$ is the {\bf scaling factor} representing the ``story'' of the universe and it is the single dynamical degree of freedom present in the system.

In the next, also supported by recent observations that seem to confirm this hypothesis,
we shall restrict our attention to the flat case,	$\ka=0$.
With this assumption the metric \eqref{FRW} becomes conformally flat.
This last fact, when $a(t)$ is given, can be manifestly seen
passing from the cosmological time $t$ to the {\bf conformal time}  $\tau$. The latter is determined by the diffeomorphism
$$
\tau(t):=\tau_0-\int^{t_0}_t \frac{1}{a(t')}\; dt',
$$
where $\tau_0$ and $t_0$ are two fixed constants.
Since this transformation is a diffeomorphism and $a$ is a field on the manifold $M$, it is clear that $a$ can be seen both as a function of $t$ or, as a function of $\tau$. For this reason, in the next, we shall indicate $a\circ\tau^{-1}(\tau)$ by 
$a(\tau)$ as it is custom for scalar fields over manifolds. 

We would like to stress that choosing $a(t)$ as the single  degree of freedom, we are at the same time selecting a fixed (free) background on which $a(t)$ evolves. 
The matter evolution could be analyzed on that background too, however,
since we restrict our attention to conformally flat spacetimes, another natural choice for the  background    would be Minkowksi spacetime $(\bM,g_\bM)$ taken with  the standard Minkowksi metric 
$$
g_\bM:= -d\tau^2 + dr^2+r^2d\bS^2(\theta,\varphi) \;,
$$
where $(r,\theta,\varphi)$ represents a point in $\bR^3$ in the standard spherical coordinates.
In order to fix the notation, in the next, we shall indicate the generic element $x$ of $\bM$ by $(\tau,{\bf x})$; notice that this defines a coordinate system also on $M$.
Later it will also be useful, to consider as a background, the static Einstein universe $(M_e,g_e)$. The advantage of this last selection  
is in the compactness of its spatial sections.
All these spacetimes are connected by a chain of conformal embeddings \cite{Pin} $\eta$ and $\eta'$ such that
\beq\label{pullback}
(M,g) \underset{\eta}{\hookrightarrow}  (\bM,g_\bM) \underset{\eta'}{\hookrightarrow} (M_e,g_e) 
\eeq
where the conformal factor of 
$\eta$ is $\Om^2:=a^{-2}$, while the one of $\eta'$ is
$$
{\Om'}^2:=\at \frac{(1+(\tau+r)^2)(1+(\tau-r)^2)}{4} \ct^{-1}\;. 
$$
We would like to stress that in every spacetime under investigation, namely for every $a(t)$, there is a conformal Killing vector, $\pa_\tau$ generating conformal translations $\tau\to\tau+\la$.
We shall use these conformal symmetry in order to define a  suitable quantum state used for the computation of backreaction.

\subsection{Scalar fields, quantization and the stress tensor}

As a model for matter we shall consider a real scalar field, namely a field satisfying
	\beq\label{KleinGordon}
	P\varphi=0 ,\qquad P= -\Box_g+\xi R_g + m^2.
	\eeq
For simplicity, we shall choose the coupling to the metric to be $\xi=1/6$, corresponding to the so-called conformal coupling, whereas we shall allow the mass to be non vanishing. In this respect $m$ shall play the role of the ``coupling constant'' for the matter gravity interaction, or the interaction between the field $\varphi$ and the scaling factor $a$.

We shall employ the quantization prescription proper of the algebraic approach (\cite{BFV,HW01}). First of all we select the $*-$algebra of local observables / fields $\mA(M)$ and afterwards, choosing a suitable state $\bom$ (a normalized, positive linear functional over $\mA(M)$), the expectation values for the elements of $\mA(M)$ are obtained.
In order to keep the paper self-consistent, an explicit construction of $\mA(M)$ will be given in the appendix \ref{deformationquantization}.
For our purpose it is important to stress that, being $M$ a globally hyperbolic spacetime, $\mA(M)$ is uniquely determined by the equation of motion and the unique causal propagator\footnote{The construction, and the proof of existence and uniqueness can be found for example in  \cite{BGP}.}  $E$.
The whole procedure is hence functorial as discussed in \cite{BFV}.
%
%
%
In order to compute the backreaction of matter on gravity by means of the semiclassical Einstein equations \nref{semiclassical1}, the expectation values of the component of the stress tensor need to be computed.
Hence, since the pointwise product of fields is singular, the quantum state has to be selected in a physically reasonable way. In other words a certain regularization procedure has to be implemented   and it must be meaningful on that state.

We proceed now to discuss this aspect in detail. On the field algebra $\mA(M)$ (constructed as discussed on the appendix \ref{deformationquantization}), the functional $\bom$ representing the state can be thought as being completely described by a suitable set of distributions $\om_n\in\cD'(M^n)$, the so called {\bf  $n-$point functions}.
In this paper we shall restrict our attention to states that are quasi-free, namely states whose $n-$point functions vanish for odd $n$ while  for even $n$ they can be given in terms of the two-point function $\om:=\om_2$ only by means of the formula
\beq\label{npoint}
\om_n(x_1,\dots, x_n) = \sum_{\pi\in P_n} \om(x_{\pi(1)},x_{\pi(2)}) \dots 
\om(x_{\pi(n-1)},x_{\pi(n)})
\eeq
where the sum is taken over the set $P_n$ of permutations of the first $n$ natural numbers with the constraints
$$
\pi_n(2i-1)<\pi_n(2i), \qquad 1\leq i\leq n/2\;  
\qquad \text{and}  \qquad
\pi_n(2i-1)<\pi_n(2i+1)\;, \qquad 1\leq i <  n/2\;.
$$
Let us proceed to discuss the regularization prescription we shall use   to give meaning to observables like the component of $T_{\mu\nu}$ or the field square.
In order to implement it in practice, we must have control on the 
 ultraviolet behavior of our states. For this reason, 
 we shall impose a constraint on the states, asking that 
 their two-point functions have singularities that look, as much as possible, similar to the ultraviolet singularity displayed by the Minkowski vacuum. 
 This is achieved requiring that the two-point function $\om$ satisfies the so called {\bf microlocal spectrum condition} \cite{Radzikowski,BFK}.
\begin{definizione} 
A state $\bom$ on $\mA(M)$ satisfies 
the microlocal spectrum condition ($\mu$SC) if its two-point function $\om$ is a distribution on $C^\infty_0(M^2)$ whose wave front set has the form
$$
WF(\om)= \{ (x,y,k_x,k_y) \in T^*M^2 \setminus \{0\}\;, \;\;(x,k_x)\sim (y,-k_y)\;,\;\; k_x \triangleright 0   \} \;.
$$
\end{definizione}
The relation $(x,k_x)\sim (y,-k_y)$ is satisfied in $M$ if the points $x$ and $y$ are connected by a null geodesic whose cotangent vector in $x$ is $k_x$ and in $y$ is $-k_y$.
Furthermore $k_x \triangleright 0$ if $k_x$ is future directed.
Notice that, thanks to the work of Sanders 
\cite{Sanders}, it suffices to give a condition only on the two-point function.
In the above mentioned paper it is actually shown that 
every state of a free scalar field algebra, whose two-point function satisfies the microlocal spectrum condition, has $n-$point functions that enjoy the microlocal spectrum condition too (for the definition in that case we refer to the work \cite{BFK}). 

Notice that, the two-point function of a quasi-free state for fields satisfying the equation of motion \nref{KleinGordon} and the microlocal spectrum condition, has an ultraviolet singularity that is similar to the one of the Minkowski vacuum.
In other words, its integral kernel 
can be expanded in the following way
\beq\label{Hadamard}
\om(x,y):=  \lim_{\epsilon\to 0^+} \at \frac{U(x,y)}{\si_\epsilon(x,y)} + V(x,y) \log \frac{\si_\epsilon(x,y)}{\la^2} \ct +W(x,y)
\eeq
where $U, V$ and $W$ are smooth functions on $M^2$, $\lambda$ is a length scale used in order to have the logarithm with a dimensionless argument and $\si_\epsilon(x,y)= \si(x,y) +i\epsilon (\mT(x)-\mT(y))+\epsilon^2$ where $\mT$ is a generic time function and $\si$ is the Synge world function, namely half of the squared geodesic distance  taken with sign \cite{DeWitt, KW}. 
Furthermore, $\epsilon$ is a regularization parameter that eventually tends to $0^+$ (weak limit).
A two-point function whose singular part is of the form \nref{Hadamard} is said to be of {\bf Hadamard form}. 
It is important to notice that the coefficient $U$ and $V$ are completely determined out of the geometry and of the equation of motion, while the coefficient $W$ really characterizes the state.
The distribution constructed out of the first two terms in \nref{Hadamard} is called {\bf Hadamard singularity} and it is indicated by $\cH$, then its integral kernel is
\beq\label{Hadamardsingularity}
\cH(x,y):= \lim_{\epsilon\to 0^+} \at  \frac{U(x,y)}{\si_\epsilon(x,y)} + V(x,y) \log \frac{\si_\epsilon(x,y)}{\la^2} \ct
\;.
\eeq
Hence, in order to regularize the states satisfying the microlocal spectrum condition we can simply subtract the singular part $\cH$ from $\om$, what remains is a smooth function whose coinciding point limit can be safely computed.
As a side remark, we notice that, since we are interested in local fields, for our purpose it is enough to have an explicit form of $\cH$ on small domains.
The question about global existence of $\cH$ is hence not an important one in our approach. 
%
%
Coming back to the main point, it is hence possible to obtain a well posed stress tensor on states satisfying the $\mu$SC. On practical ground this is achieved by operating on the regularized two-point function $\om-\cH$  with a certain differential operator before computing the well posed coinciding point limit.
This procedure is called {\bf point splitting regularization}.
Furthermore, we notice that on those states 
the fluctuations of the observables computed on similar states are always finite.
%

Let us discuss this point in some details.
On the class of states that satisfy the $\mu$SC, we could extend the algebra of fields $\mA(M)$ in order to encompass also pointwise product of fields.
Here we shall indicate by $\mF(M)$, the extended algebra of fields, namely the $*-$algebra that contains also the Wick powers as constructed in \cite{HW01}, or more recently by means of deformation quantization methods in \cite{BDF,BF}. For completeness, the main steps of that construction can be found in the appendix \ref{deformationquantization}.
We are now ready to discuss the form of the stress tensor as an element of $\mF(M)$. To this end, following the discussion introduced in \cite{Mo03}, we consider the following operation on smooth field configurations $\varphi$
$$
T_{ab}(\varphi):=\pa_a \varphi \pa_b \varphi - \frac{1}{6}g_{ab}\at \pa_c\varphi\pa^c\varphi
+m^2\varphi^2 \ct
-\xi \nabla_a\pa_b \varphi^2
+\xi\at  R_{ab}-\frac{R}{6} g_{ab} \ct \varphi^2
+\at\xi-\frac{1}{6}\ct g_{ab} \Box \varphi^2.
$$
Notice that for every compactly supported smooth function $f$,
$$
T_{ab}(\varphi)(f):= \frac{1}{2} \langle f,T_{ab}(\varphi)  \rangle 
$$
is an element of $\mF(M)$, and 
we shall use it in order to compute the backreaction in \eqref{semiclassical1}.
Another element of $\mF(M)$ that will be useful is $\phi^2(f)$. This 
is nothing but $\int f \varphi^2/2$.
We shall indicate by $\langle \phi^2  \rangle_{\bom}$ and $\langle T_{ab}  \rangle_{\bom}$ the expectation values of the two observables introduced above computed on the state $\bom$ regularized according to the previous discussion.
Hence $\langle \phi^2  \rangle_{\bom}$ and $\langle T_{ab}  \rangle_{\bom}$ are obtained replacing the smooth field configuration $\varphi(x)\varphi(y)$ with the smooth function $\om-\cH$ in $T_{ab}(\varphi)(f)$ and $\phi^2(f)$ before performing the coinciding point limit (contracting with the delta function).
Further details on this regularization, can be found in appendix \ref{deformationquantization}.
In the procedure presented above, there is an ambiguity in the definition of $\cH$, whose form is described in $\cite{HW01}$; we call it renormalization freedom, later on we shall give more details on this crucial aspect.

\se{Homogeneous states and Hadamard property in flat FRW}

\subsection{Characterization of pure, homogeneous states in terms of solutions of the wave equation.}

In this subsection we shall proceed  to discuss the form of the states we are interested in, we shall make use of some previous works in order to get the form of the two-point function of the pure homogeneous  states on FRW spacetimes in particular.
We start with the definition of the class $\mC(M)$ of pure and homogeneous quasi-free states, we shall employ in the present paper.
Although the set $\mC(M)$ does not contain all the possible pure and homogeneous states for $\mA(M)$, 
that definition is motivated by the works
 of  L\"uders and Roberts \cite{LuRo} (Theorem 2.3 of \cite{LuRo} in particular) and the one of Olbermann \cite{Olbermann}
see also the nice work of Degner and Verch \cite{DV} for a review.
Furthermore, we shall restrict our attention to that class.

\begin{definizione}\label{statiomogenei}
We shall indicate by $\mC(M)$ the set  of {\bf pure}, {\bf homogeneous} and quasi-free states $\bom$ for $\mA(M)$ whose two-point function $\om\in \cD'(M\times M)$, using the standard Minkowksian coordinates, takes the following form, in the distributional sense\footnote{Here we mean that $\om$ has to be constructed as the weak limit $\epsilon\to 0^+$ of the expression on the right hand side of \eqref{twopointT} regularized  multiplying  the integrand by an extra factor $e^{-\epsilon |{\bf k}|}$.},  
\beq\label{twopointT}
{\omega}((\tau,{\bf x});(\tau',{\bf x}')):= \frac{1}{(2\pi)^3} \int_{\bR^3}   \frac{\overline{S_k}(\tau)}{a(\tau)}\frac{S_k(\tau')}{a(\tau')} e^{i{\bf k}\cdot({\bf x}-{\bf y})}     d{\bf k} \;,
\eeq
where $(\tau,{\bf x})$ is a generic element of $(I\times \bR^3,g)$,
 $k= |\bf k|$. For every $k\geq 0$, $S_k$ is a smooth function 
 satisfying 
\beq\label{eqmotoT}
\frac{d^2}{d\tau^2}S_k(\tau)+(m^2 a(\tau)^2 +k^2)S_k(\tau)=0,
\eeq
and 
\beq\label{Wronskian}
\overline{S_k}(\tau) \frac{d}{d\tau} S_k(\tau) - 
\frac{d}{d\tau} \overline{S_k}(\tau) S_k(\tau) =i \;.
\eeq
Furthermore, for every $\tau$, the map $k\mapsto S_k(\tau)$, and its first time derivative, are measurable functions that are polynomially bounded at large $k$. When restricted on a generic interval $[0,k_0] \subset \bR^+$ they are in $L^2([0,k_0], k^2 dk)$.
\end{definizione}

It is not guaranteed that the state, constructed out of a generic solution $S_k$ of the equation \nref{eqmotoT} (also with the constraint on the Wronskian), is of Hadamard type.
It is also usually not easy to impose some ad hoc initial conditions for $S_k(\tau)$ on a spacelike hypersurface in order to have an Hadamard state.
This problem, in connection with the concept of adiabatic states, has been discussed in the paper of Junker and Schrohe \cite{JuSc} who showed that it is possible to get Hadamard states only if the recursive construction of adiabatic vacua does not break down. However, as shown by L\"uders and Roberts \cite{LuRo}, one cannot exclude such possibility.
Another possible explicit construction of Hadamard states is the one presented by Olbermann \cite{Olbermann}, but this construction requires the smearing on a finite time interval. It is hence very difficult to use similar ideas in treating backreaction problems.
Here we would like to give an alternative construction employing ideas similar to the one presented 
in \cite{Moretti,DMP2,DMP4} and adjusting them in order to be suitable for the purpose of studying the gravity-matter interacting  system.
First of all, we start selecting a particular solution $\chi_k(\tau)$ of \nref{eqmotoT}, when the spacetime $M$ admits the limit $\tau\to-\infty$, namely the one that satisfies the initial conditions 
\beq\label{initialconditionchi}
\lim_{\tau\to -\infty}e^{ik\tau}\chi_k(\tau)  = \frac{1}{\sqrt{2k}}\;,\qquad
\lim_{\tau\to -\infty}e^{ik\tau}\frac{d}{d\tau}\chi_k(\tau)  = -i \sqrt{\frac{k}{2}}\;.
\eeq
We can now introduce the following proposition that summarizes some results obtained in \cite{LuRo,Olbermann,DV}

\begin{proposizione}
If the limits in \nref{initialconditions} are well posed, every solution $S_k$ of \nref{eqmotoT} which determines a state $\bom\in\mC(M)$ according to the definition \ref{statiomogenei} can be  realized as a linear combination of the $\chi_k$ (the solution of the equation \nref{eqmotoT} satisfying the asymptotic conditions \nref{initialconditionchi}):
\beq\label{defT}
S_k(\tau):=A(k) \chi_k(\tau) +B(k) \overline \chi_k(\tau),
\eeq
and the coefficient $A(k)$ and $B(k)$ are such that 
$$
|A(k)|^2 - |B(k)|^2 =1\;.
$$
Furthermore, for every $\tau$, the functions, on $\bR^+$, $k\mapsto A(k)\chi_k(\tau)$ and $k\mapsto B(k)\chi_k(\tau)$, together with 
$k\mapsto A(k)\frac{d}{d\tau}\chi_k(\tau)$ and $k\mapsto B(k)\frac{d}{d\tau}\chi_k(\tau)$, are at most of polynomial growth and locally square integrable.
\end{proposizione}

Notice that, even if we restrict our attention to the class of pure states\footnote{More details in this respect can be found in \cite{LuRo,Olbermann} },
we have still some freedom in the definition of the state, as expressed in the preceding proposition by the choice of $A(k)$ and $B(k)$ in the $S_k$ of \nref{defT} used in finding the two-point function  \nref{twopointT}. Motivated by this expression, we shall indicate by $\bom_{A,B}$ the quasi-free state whose two-point function  is 
\beq\label{omegaAB}
\omega_{A,B}((\tau,{\bf x});(\tau',{\bf x}')):= \frac{1}{(2\pi)^3}\int_{\bR^3} d{\bf k}\;  e^{i{\bf k}({\bf x} -{\bf x}')} \frac{\overline{S}_k(\tau)}{a(\tau)} \frac{S_k(\tau')}{a(\tau')} 
\eeq
where we have highlighted the dependance upon $A(k)$ and $B(k)$ in the state.

We are now going to find sufficient conditions that have to be fulfilled by 
$A(k)$ and $B(k)$ in order to ensure that $\bom_{A,B}\in\mC(M)$  satisfies the Hadamard condition.
The strategy we shall pursue is the following one. We prove that $\om_{1,0}$ is of Hadamard form provided some regularity is shown by $a(t)$ in \nref{FRW} and then that 
$\bom_{A,B}$ satisfies the $\mu$SC if and only if $B(k)$ is rapidly decreasing for large $k$.
Notice that, in order to prove that a certain distribution satisfies  the microlocal spectrum condition, it is not enough to test it on some Cauchy surface because no remnant of causality survives the projection of any vectors on a spacelike hypersurface.  This is not true for projection on lightlike hypersurfaces, and such is the reason why the arguments used in \cite{Moretti}, \cite{Hollands} and  \cite{DMP4} works.
We would like to give sufficient conditions on the form of the spacetime, and on its past asymptotic form in particular, in order to guarantee that $\om_{1,0}$ is of Hadamard form.
In the case of asymptotic de Sitter spacetime, it was already shown in \cite{DMP4} that the state $\bom_{1,0}$ satisfies the microlocal spectrum condition, and that if the spacetime is precisely the de Sitter one, $\bom_{1,0}$  becomes the well known Bunch Davies state \cite{BD}.
Furthermore, in the case of asymptotic flat spacetimes, Moretti has proven the Hadamard property for an analogue state \cite{Moretti}; a similar proof can also be found in \cite{Hollands}.
We are now going to generalize these results in order to encompass the case considered here. The idea we would like to use, is to consider $\om_{1,0}$ as the pullback $(\eta\circ\eta')^*(\tilde{\om})$ associated with the  conformal embedding $\eta\circ\eta'$ introduced in \eqref{pullback} where $\tilde\om$ is the two-point function of a certain state in a hyperbolic subset of the Einstein universe that contains the image of $M$ under $\eta'\circ\eta$ together with its past boundary $\scrim\cup i^-$.
Notice that on the latter, the field theory looks like a massless Klein Gordon field perturbed with an external potential of the form $m^2 a(\tau)^2 (1+(\tau+r)^2)(1+(\tau-r)^2)/4$, and hence $\tilde{\bom}$ is a state for this particular theory. 
It happens that, if that potential becomes regular on the extended spacetime in a neighborhood of $\scrim \cup i^-$ then the state $\bom_{1,0}$, defined in such a way, turns out to be of Hadamard form.

\begin{teorema}\label{HadamardTheorem}
Suppose that $M=I\times \bR^3$ where $I=(0,t_0)$ and that $a(t)$ is such that the region $t\to 0$ corresponds to the past null infinity of $(\bM,g_{\bM})$ in $(M_e,g_e)$ and furthermore such that the smooth function
\beq\label{masspotential}
(1+(\tau+r)^2)(1+(\tau-r)^2)m^2a(\tau)^2
\eeq
defined on $M$ seen as embedded in the Einstein universe $(M_e,g_e)$  can be smoothly extended  in a neighborhood $\cO\subset M_e$ containing the past boundary of $M$ here indicated as $\scrim \cup i^-$.
Under these hypotheses the state $\bom_{1,0}$, constructed out of the 
solutions $\chi_k$ that satisfy the initial condition on $\Im^-$ satisfies the $\mu$SC.
\end{teorema}

\begin{proof}
The proof can be found following similar ideas as those presented in \cite{Moretti, Hollands}
or alternatively in \cite{DMP4}, we shall summarize here the main strategy and we shall discuss only the key point in detail.
We would like to show that the two-point function $\om_{1,0}$  satisfies the microlocal spectrum condition, hence we have to analyze its wave front set and to show that
\beq\label{muSC}
WF(\om_{1,0}) = 
 \left\{ (x,y,k_x,k_y) \in T^*M^2 \setminus \{0\}\;, \;\;(x,k_x)\sim (y,-k_y)\;,\;\; k_x \triangleright 0   \right\} \;.
\eeq
Notice that, in showing the preceding equality, the difficult part is to prove the inclusion $\subset$. If it holds true, the other inclusion descends from the fact that $\bom_{1,0}$ is a state  \cite{SV01,SVW02} and out of an application of the H\"ormander's theorem of  propagation of singularity \cite{DH}. 
In order to show the validity of inclusion $\subset$ we can rewrite 
$\om_{1,0}(f,g)$ as    $\mathfrak{w} (\Om E f\rest_\scrim   , \Om E g\rest_\scrim)$ where $\mathfrak{w}$ is a certain distribution on $\scrim \times \scrim$
defined employing the Fourier Plancherel transform\footnote{This operation is introduced and discussed in the appendix of \cite{Moretti}.} along the $\tau-$coordinate on $\scrim$:
$$
\mathfrak{w}(\psi_1,\psi_2):= \int_{\bS^2}\int_0^\infty 2k\;  \overline{\widehat{\psi_1}(k,\theta)}  \widehat{\psi_2}(k,\theta)\;
dk  d\bS^2 \;,
$$
where $d\bS^2$ is the measure on the two dimensional unit sphere $\bS^2$ and $\theta$ is a shortcut for the standard spherical coordinates.
A discussion on such a distribution can be found in theorem 3.1 of \cite{Moretti}.
Furthermore, $\Om E f \rest \scrim$ is the restriction to $\scrim$ of the wave function $E(f)$ constructed by means of the causal propagator $E$, multiplied by a certain factor $\Om$.
That restriction can be safely computed in an Einstein universe and the multiplication by the factor $\Om$ corresponds to the conformal transformations that maps solutions of $P$ in $M$ to solutions of $P_e$ in $M_e$.
We have indicated by $P_e$  the operator $-\Box_e + \frac{1}{6} R_e + m^2 a^2 \Om^2  $ where $\Box_e$ is the d'Alembert operator associated with the Einstein metric.
Hence, the two-point function $\om_{1,0}$ can be seen as a composition of distributions 
$$
\om_{1,0} :=     \mathfrak{w} \circ \at \Om E \rest_\scrim \otimes \Om E\rest_\scrim \ct\;.
$$
Because of the non compact nature of the restriction of $E$ on $\scrim$, which persists also if the other entry is localized on a compact set of $M$, the previous composition of distributions cannot be straightforwardly justified  by an application of theorem 8.2.13 of \cite{Hormander}. 
On the other hand, in proposition 4.3 in \cite{Moretti}, it is proven that a similar composition is well defined 
and that the result has the desired wave front set
provided the following two facts hold:
First of all considering a compact set $\cO$ inside $M$, there exists a neighborhood $\cO_{i^-}$ of $i^-$ in $M_e$ such that there are no lightlike geodesics connecting any point in $\cO$ with those in $\cO_{i^-}$, and, second, if one entry of the causal propagator is restricted on $\cO$ and the other on $\scrim$, the causal propagator itself  must fall off sufficiently fast towards $i^-$ in $M$.
Both requirements hold under the hypotheses of the present theorem, 
the first descends from the fact that $M$ is conformally flat and the latter is a consequence of the fact that $m^2 a(\tau)^2 (1+(\tau+r)^2) (1+(\tau-r)^2)$ can be smoothly extended on a neighborhood of $i^-$ in $M_e$, hence the same result as the one stated in lemma 4.2 of \cite{Moretti} holds in the present case too. 
\end{proof}


\vsp
\noindent
Notice that the constraints given above are only sufficient but not necessary, in fact in \cite{DMP4} it has been proven that $\bom_{1,0}$ is a well defined Hadamard state also when the spacetime is asymptotically de Sitter in the past, and in such a case the potential that appear in the equation of motion on the Einstein universe is singular on $i^-$. 
Hence the hypothesis 
 of smooth extension of the function \eqref{masspotential} over $\scrim\cup i^-$ in $M_e$ can be relaxed. 
Finally, notice that employing similar techniques to the one presented in \cite{DMP2} and in \cite{DMP4} based on a careful analysis of the modes $\chi_k(\tau)$, it is actually possible to obtain the Hadamard property for the state $\bom_{1,0}$ also when \eqref{masspotential} can be extended over $\scrim\cup i^-$ only with the regularity of $C^1(M_e)$.
We would like to conclude this section by introducing a theorem that gives some necessary and sufficient condition for the state $\bom_{A,B}$ to be of Hadamard form.

\begin{teorema}\label{differenza}
Suppose that $\om_{1,0}$, constructed as in \eqref{omegaAB}, is of Hadamard form. Then the quasi-free state
$\bom_{A,B}\in\mC(M)$, satisfies the microlocal spectrum condition if and only if $|B(k)|$ decreases rapidly for large $k$, in other words, there are constants $c_n$ and $\underline{k}$ such that 
$$
|k|^n |B(k)| \leq c_n\;, \qquad   \forall\; |k| > \underline{k} \;.
$$
\end{teorema}

\begin{proof}
The proof of this fact can be easily done using the formulation of the state on a given Cauchy surface, see \cite{LuRo} for a discussion.
Notice that, when restricted on a Cauchy surface, the difference between the two-point functions $\om_{A,B}-\om_{1,0}$ becomes smooth if and only if $B(k)$ decreases rapidly (and this holds for the time derivative of one or both of its entries too).
This proves that the rapid decrease of $|B(k)|$ is a necessary condition for the validity of $\mu$SC. 
Finally, in order to prove that it is also sufficient, we can make use of a formula like the one given in equation 3.60 of \cite{KW} and reconstruct $\om_{A,B}-\om_{1,0}$ out of its form on a Cauchy surface.
We can extend the result about the smoothness to the whole $M^2$ by noticing that every of the four terms appearing in the formula 3.60, above cited, can be obtained composing $\om_{A,B}-\om_{1,0}$ (sometime with a time derivative applied to one or both of its entries) 
 with the tensor product $E\otimes E$ (sometime taken with a time derivative applied to one or both of its entries).
The proof can be easily finished noticing that $E$ maps compactly supported smooth functions on the Cauchy surface to smooth functions in the spacetime.
\end{proof}

\se{Semiclassical treatment of backreaction and existence of solutions}

\subsection{Semiclassical approximation}

We shall now proceed to treat the backreaction of quantum matter on gravity, and, as discussed in the introduction, we shall do it for flat FRW spacetimes.  
With such a large symmetry, the Einstein equations take the well known  Friedmann Robertson Walker form, namely, being the stress tensor of matter diagonal and equal to ${{T_{\mu}}^\nu}=diag(-\rho,P,P,P)$, if we indicate by $\dot{a}$ the derivative of $a$ with respect to the cosmological time $t$ and $H:=a^{-1} \dot{a}$ they become 
$$
3H=8\pi \rho - \frac{3\ka}{a^2} \;,    \qquad    3\at \dot{H}+H^2 \ct=-4\pi(\rho+3P) \;.
$$
Because of the conservation of the stress tensor we can instead solve the  equation for the trace of $G_{\mu\nu}$  and of $T_{\mu\nu}$
$$
-R=8\pi T
$$
and then employ the first FRW equation as a constraint, that needs  to be imposed only at a fixed time, in order to fix only an initial condition.
We shall assume this point of view and, in the semiclassical scenario, we have simply to substitute the classical $T$ with the expectation value of the corresponding  observable  $\langle T\rangle$ computed in the chosen quantum state. 
More precisely, as discussed in the introduction, we have to select a class of states (one for every $a(t)$) and then
the semiclassical equation $-R=8\pi \langle T\rangle $ acts as a constraint on this class.
Notice that, by employing the initial values on the null boundary, we have already disentangled the problem of the definition of the class of states from the geometric property of the spacetime.

Hence the backreaction of such system can be treated simply employing the equation for the trace. But in the trace, taking the conformal coupling and a non-vanishing mass, the explicit form of the state become important.
More precisely, the expectation value of the trace of the stress tensor in a state $\bom$ is
$$
\langle {T}\rangle_{\bom}:=\frac{2[V_1]}{8\pi^2}+\at-3\at\frac{1}{6}-\xi\ct\Box -m^2\ct    \langle\phi^2\rangle_\bom.
$$
Above $[V_1]$ is the coinciding point limit of the second Hadamard coefficient (see \cite{SV01, Mo03, DFP} for more details).
This contribution to the trace is the anomalous one, it coincides with the well known trace anomaly displayed by the stress tensor \cite{Wald78} in the massless case.
Coming back to the dynamical equations, in the case of a flat FRW metric, the equation of motion for the scaling factor is 
\beq
\label{semiclassical}
-6\at\dot{H}+2 H^2\ct=
-8\pi  m^2 \langle\phi^2\rangle_\bom
-\frac{1}{30\pi}\at\dot{H}H^2+H^4 \ct 
\;.
\eeq
In writing the preceding equation we have performed some choices of the renormalization parameters, in particular we have chosen some of them in such a way that the higher derivatives of the coefficients of the metric disappear.
There is an extra renormalization freedom that we have not considered here, but it can always be reabsorbed in the freedom present in $\langle\phi^2\rangle_\bom$.
It seems important to notice that, apart from these renormalization freedom, the trace anomaly is fixed and it is really a $c-$number.
For the discussion of these and other points we recall the work \cite{DFP}.

Hence, in the case of vanishing $m$, the state does not enter anymore in \eqref{semiclassical} and an exact solution can be easily found\footnote{see \cite{Wald78a, DFP} for more details}.
This solution can be written in an implicit form as
\beq\label{massless-solution}
e^{4 t H_+} = e^{ \frac{2 H_+}{H}} \left| \frac{H+H_+}{H-H_+}   \right|
\eeq
where the time of the big bang is fixed to be  $t=0$  and $H_+^2 = 360 \pi $ in natural units, that is $G=\hbar=c= 1$.

The massless solution \eqref{massless-solution} presents three branches for $H>0$ (the first $0<H^2<H_+^2/2$, the second $H_+^2/2<H^2<H_+^2$ and the last $H^2> H_+^2$), we concentrate ourself on the upper one and we discuss some of its properties.
Notice that, it corresponds to a universe that shows, at the beginning, a phase typical of the power law inflationary scenario, while it ends up in a phase which corresponds to a de Sitter universe.
It is a remarkable fact that the form of the initial singularity is  lightlike; hence, for example, the horizon problem is not present in this simple model.
The Hubble constant $H$ of the de Sitter phase shown in the asymptotic future is equal to $H_+$, and it is many order of magnitude too big in order to explain the present acceleration of the universe.
Despite of this fact, in \cite{DFP}, it has been shown that, in the case of massive fields, this parameter is not fixed and it needs to be interpreted as an extra renormalization freedom.
In the next we shall make use of the lightlike nature of the initial singularity to see if it is possible to obtain an exact solution also in the case of massive fields, and we shall look for solutions that are similar to the massless one at the beginning of the universe. In this respect, we shall treat $H_+$ as a renormalization constant.
Furthermore, since, in the massless case, only the anomaly appears in the trace and it is a $c-$number it has a vanishing variance, hence,
such equation should hold also for large $H$ (as is the case close to the initial singularity $t=0$ in the \eqref{massless-solution}).
Of course, the situation changes when the mass is different from zero, and in this case the result provided by the semiclassical Einstein equations needs to be interpreted more carefully.
We shall comment on this points in the next section.

\subsection{Initial conditions at the beginning of the universe}

In this subsection we would like to discuss the existence of solutions of the semiclassical Einstein equations \eqref{semiclassical} that look as much as possible close to \eqref{massless-solution}, at least on small intervals of proper time just after the initial singularity.
Hence, we shall assume the form of the singularity to be lightlike just as in the massless case \eqref{massless-solution}.
As previously discussed this task can be accomplished only after a careful analysis of the class of states we have to use in order to obtain the expectation values of $\phi^2$ in the different spacetimes we are considering, in any case the requirement of having a lightlike initial singularity is essential in order to apply the results of the preceding sections and to have 
Hadamard states.

The states we would like to chose are the one constructed in the preceding part of this paper and denoted by $\bom_{A,B}$ with $B$ of rapid decrease. 
We shall start considering only the state $\bom_{1,0}$, and we notice that these states are defined by the requirement to be vacuum states with respect to the conformal time on the initial singularity
(this procedure works only for lightlike initial singularities).
Before proceeding with the discussion of equation \eqref{semiclassical} we have to discuss the form of the expectation value of $\phi^2$ on these states and to fix once and forever the renormalization freedom, according to the picture presented in \cite{BFV}.

Since we would like to impose initial conditions on $\scrim$ or better at $\tau\to-\infty$, where the Hubble constant $H$ usually diverges,
it is useful the rewrite the dynamical equation for its inverse that we shall call $X:=H^{-1}$. With this in mind, we can rewrite the equation \eqref{semiclassical} as follows
\beq\label{semiclassical2}
\frac{d X}{dt} =
1-
\frac{H_c^2 \; X^2 }{1-H_c^2 \; X^2} 
+ 
\frac{C_m^2 \; X^4 }{1-
H_c^2 \; X^2}\langle\phi^2\rangle_{\bom_{1,0}}\;,
\eeq
where $H_c^2:=H_+^2/2=180\pi$ and $C_m^2 := 240\pi^2m^2 $ are two constants.
Notice that, since $\langle \phi^2 \rangle_{\bom_{1,0}}$ is a functional of $a(t)$ and of $\tau$, the proof of the existence of solutions of \eqref{semiclassical} needs some effort, and the analysis of the functional dependance of $\langle\phi^2\rangle_{\bom_{1,0}}$ upon $X=H^{-1}$ needs to be carefully performed in particular.

We intend to prove the existence of solutions of \eqref{semiclassical} (or better of \eqref{semiclassical2})
in a certain compact set $\cB_c$ of a Banach space $\mB$. To this end let us start introducing these spaces of functions.
\begin{definizione}\label{Bc}
Consider the set $\mB$ as the subset of elements of $C^1([0,t_0])$ that vanish in $0$ and whose first derivative is bounded on $[0,t_0]$.
$\mB$ is a Banach space when equipped with the norm
$$
\| f \|_\mB:= \sup_{[0,t_0]} \left| \dot{f}(t)\,  \right|\;.
$$
For every $\frac{2}{3}<c<1$ we define $\cB_c$ as the closed ball in $\mB$ of radius $1/2-c/2$ centered in 
\beq\label{center}
f_0(t):= \frac{{1+c}}{2}\; t\; \qquad (t\in [0, t_0]).
\eeq
\end{definizione}

The strategy of the proof of existence of solutions of \eqref{semiclassical2} is the following: 
We shall build a sequence $X_n$ in $\cB_c$ by a recursive use of a certain map $\cT:\cB_c\to\mB$ defined as 
\beq\label{contraction}
\cT\at X \ct(t) := \int_{0}^t \aq \frac{1-2 H_c^2\; X^2 }{1- H_c^2 \; X^2} 
+ \frac{C_m^2\; X^4}{1-{H_c^2\; X^2}}\langle\phi^2\rangle_{\bom_{1,0}}[X] \cq dt'
\eeq
 and we shall show that, since $\cT$ it is contractive on $\cB_c$, $X_n$ converges to a solution. Hence the existence and uniqueness descends from an easy application of the Banach fixed point theorem.
Notice that, if we succeed in obtaining that result, since certain physical properties (like the presence of a phase of power law inflation) are enjoyed by every element of $\cB_c$ those properties have to be shown by the solution too.

We shall now discuss how it is possible to associate a spacetime to an $X\in\cB_c$ in such a way that the initial conditions 
 are implemented.

\begin{proposizione}\label{functorialspacetime}
Fix the three constant $t_0$, $a_0$ and $\tau_0$. Then every element $X$ of $\cB_c$ determines a spacetime $(M,g[X])$ where $M=[0,t_0]\times \bR^3$
and where $g[X]$ is the metric \eqref{FRW} constructed out of the following scaling factor 
$$
a[X](t):= a_0 \exp\at -\int^{t_0}_{t} X(t')^{-1}  dt' \ct\;.
$$
The Hubble parameter shown by $(M,g[X])$ is then 
$$
H(t)= X(t)^{-1}\;,
$$
and the corresponding conformal time
$$
\tau[X](t):=\tau_0-\int_{t}^{t_0} a[X](t')^{-1} dt' \;,
$$
is a functional of $X$ too.
Finally, the spacetime enjoys the following initial conditions
\beq\label{initialconditions}
X(0)=0 \; ,\qquad a(t_0) = a_0, \qquad \tau(t_0)=\tau_0\;
\eeq
per construction.
\end{proposizione}


For every $X$ in $\cB_c$ we have immediately, from the definition, that the initial conditions \eqref{initialconditions} given above are satisfied, and 
the functional $a[H]$ and $\tau[H]$ are, moreover, well defined on $\cB_c$.
Notice that, on $\cB_c$ 
$$
\lim_{t\to 0 } \int^{t_0}_t X^{-1}(t') dt' = \infty 
$$
out of this, we have that  $a(0)=0$ sufficiently fast in order to ensure that  $\tau[X]\to -\infty$ for $t\to0$.
This last requirement is necessary in order for the resulting spacetime to have  a lightlike past boundary or, in other words, that the initial surface $t=0$ is of null type. As already discussed, this last requirement is a necessary prerequisite for the construction of Hadamard states on 
$(M,g)$. We have in fact, the following two propositions
\begin{proposizione}\label{HadamardinBc}
For every spacetime characterized by a metric of FRW type, whose scaling factor $a(t)$ is such that $X(t)=\frac{a(t)}{\dot{a}(t)}$ is in $\cB_c \cap C^\infty([0,t_0])$, we can construct $\om_{1,0}$ as in \eqref{omegaAB} and it determines a quasi-free state $\bom_{1,0}$ which, moreover, satisfies the microlocal spectrum condition.
\end{proposizione}
\begin{proof}
If the analogue of function \eqref{masspotential} can be smoothly extended over $\scrim\cup i^-$ the proof descends from theorem \ref{HadamardTheorem}, otherwise, also according to the discussion given after that theorem we can adapt the proof of theorem 3.1 given in  \cite{DMP4} to the present case.
%
%
\end{proof}
Notice that, if we drop the requirement of being smooth, and we replace it only with $X \in \cB_c$, the construction of the state is still valid, and, even if the resulting two-point function is not of Hadamard form, we can still apply the procedure to compute expectation values of $\phi^2$. Unfortunately, for fields that contain derivatives, it is not anymore guaranteed that the point splitting regularization works for these states (they are adiabatic states of suitable order, in the sense given in \cite{JuSc}).

We shall conclude 
this subsection by noticing that, since $a[X]$ is positive, $\tau[X](t)$ can always be inverted, furthermore, on $\cB_c$ some useful inequalities hold. We shall summarize them in the next proposition:
\begin{proposizione}\label{someinequalities}
The following inequalities hold for every $X$ in $\cB_c$ with $2/3< c < 1$ and for $t\in [0,t_0]$
  $$ c \leq \frac{X(t)}{t} \leq 1 , 
  \qquad  
  1\leq \frac{t}{X(t)} \leq \frac{1}{c} ,  
  \qquad
   \at\frac{t}{t_0}\ct^\frac{1}{c} \leq \frac{a[X](t)}{a_0} \leq \frac{t}{t_0}\; .   $$
\end{proposizione}  
The proof descends from a straightforward application of the definition of $a$ and $\tau$ out of $H$ and from the bounds satisfied by the 
elements of $\cB_c$.

\subsection{Expectation value of $\phi^2$ on the states $\bom_{1,0}$, and further renormalization freedom.}

We are now in place to discuss the form 
$\langle\phi^2\rangle_{\bom_{1,0}}$ constructed out of point splitting regularization on the quasi-free state $\bom_{1,0}$ introduced above in
\eqref{omegaAB}, namely we can construct it by 
\beq\label{regularization}
\langle\phi^2(x)\rangle_{\bom_{1,0}}
=
\lim_{y\to x}\aq\om_{1,0}(x,y) - \cH(x,y)\cq  + \alpha R(x) + \beta m^2
\eeq
where $\cH$ is the Hadamard parametrix \eqref{Hadamardsingularity} constructed in the spacetime 
$(M,g)$, and $\alpha$ and $\beta$ are renormalization constants.
At this point it is important to stress that $\alpha$ and $\beta$ are really constants, in the sense that they do not depend on the form of the spacetime.
In other words, if they can be fixed in a specific spacetime, they are fixed on all of them. This is a consequence of the general covariance presented in \cite{BFV} which uses the hypothesis that the fields transform covariantly under isometries used in combination with spacetime  deformation arguments.

According to the discussion presented in \cite{DFP} we would like to have Minkowksi spacetime as a solution of the semiclassical 
equation \eqref{semiclassical}, when the Minkowksi vacuum is  employed. This is tantamount to require that 
 the expectation value of $\phi^2$ on the Minkowksi vacuum in Minkowksi spacetime vanishes and this fixes the renormalization constant $\beta$ to a certain value. 
We notice also that the other renormalization constant $\alpha$ can be used in order to adjust the parameter $H_c$ present in the equation \eqref{semiclassical2}, hence from now on we shall consider it as a renormalization constant \cite{DFP}.

Once we have a prescription to fix the renormalization freedom 
we can start to discuss the Hadamard regularization; to this end we need to know the Hadamard parametrix $\cH$ in some details.
For our purposes, since the modes $\chi$ are easily treated when analyzed on the conformal Minkowksi background, 
it would be desirable to perform the subtraction $\om_{1,0}-\cH$ directly on Minkowksi spacetime.
To this avail we have to analyze how $\om_{1,0}-\cH$ transforms under the push forward $\eta_*$ associated to the conformal transformation 
$
\eta:(M,g) \to (\bM, g_{\bM})
$ 
introduced in \eqref{pullback}.
Notice that $\eta_*$ maps $\om_{1,0}$ to $a(t_1) a(t_2) \om_{1,0}$ seen as a distribution on Minkowski spacetime, and, moreover,
since the conformal transformation preserves the Hadamard property \cite{Pin}, also $a(t_1) a(t_2) \om_{1,0}$ satisfies the $\mu$SC in $\bM$.
We can hence regularize it by means of the Hadamard parametrix $\cH_\bM$ constructed in Minkowksi spacetime for the field theory $P_\bM \phi=0$ with $P_{\bM}:=-\Box_\bM + a^2m^2$.
Notice that, under the conformal transformation $\eta$, the mass term on $M$ becomes a position dependent potential $a(t)^2 m^2$.

In this respect, we already know that the microlocal spectrum condition, together with the causal propagator, are preserved by a conformal transformation
\cite{Pin}, and this holds also if the conformally coupled quantum field has a mass.
Of course, under a conformal transformation, the symmetric part of the Hadamard function is not preserved, hence
the coinciding point limit of the difference does not vanish in general, actually, taking the same scale $\la$ in the logarithmic divergence, we obtain that 
$$
\lim_{x_1\to x_2} \aq \cH(x_1,x_2) -  \frac{1}{a(\tau_1) a(\tau_2)} \cH_\bM(x_1,x_2) \cq = \frac{m^2}{8\pi^2}\log {a(\tau_2)} +  c R(x_2)\;,
$$
where $c$ is a fixed constant. 
Notice that an unexpected logarithmic dependance in the scaling factor
$\log(a(\tau))$ appears and we have to carefully use it.
On the other end, the term proportional to $R$, which appears when the two different regularization schemes are used, falls into the class of the standard regularization freedom \cite{HW01} and, thanks to the previous discussion we can incorporate it in the definition of $H_c$ and forget it.

Furthermore, since we are only interested in computing the expectation value of $\phi^2$, it is possible to regularize by subtracting an approximated version of the Hadamard parametrix constructed just out of the first Hadamard coefficient (see \cite{SV01, Mo03} for more details) 
$$
\cH^0_\bM:=\lim_{\epsilon\to0} \aq \frac{1}{8\pi^2}  \frac{1}{\si_\epsilon} + V_0\log \frac{\si_\epsilon}{\la^2}\;\cq.
$$
According to the analysis performed in chapter 4 of \cite{Friedlander}, the $V_0(x,x')$ for spacelike separated $x$ and $x'$ can be found as the result of the following integral
$$
V_0(x,x'):=\frac{1}{2} {U(x,x')} \int_0^1 \aq\frac{P_\bM U}{U}\cq(x,\ga(s))ds
$$
where the integration is done in the affine parameter $s$ of the geodesic $\ga$ connecting $x$ and $x'$ with $\ga(0)=x$ and $\ga(1)=x'$. The previous integral can be easily performed on  Minkwoksi spacetime where $U=1/(8\pi^2)$, $P_\bM= -\Box_\bM + m^2 a^2$, and the geodesics are simply the straight lines connecting $x$ and $x'$. Furthermore, if we choose $x$ and $x'$ as $(\tau,{\bf x})$ and $(\tau,{\bf x}')$ in the Minkowksian coordinates, we obtain 
$$
V_0(x,x')=  \frac{1}{(4\pi)^2}m^2 a(\tau)^2 \;.
$$
We can now write the Hadamard parametrix at first order, for $x$ and $x'$ lying on the same Cauchy surface $\Si_\tau$, as
$$
\cH^0_\bM(x,x'):= 
\frac{1}{(4\pi)^2} \at\frac{2}{\si_\epsilon} + m^2a(\tau)^2\log \at\frac{\si_\epsilon}{\la^2} \ct   \ct   \;.
$$
Notice that the first two contributions are the same as the one of a massive Klein Gordon field propagating in Minkowski spacetime with the mass equal to $m^2a^2(\tau)$ (where here $\tau$ is fixed by the choice of $\Si_\tau$). Furthermore,
we know that in this case, the coinciding point limit of the following 
expression yields
$$
\lim_{x'\to x}
\left[ \cH^0_\bM(x,x') -  \frac{1}{(2\pi)^3} \int \frac{1}{2\sqrt{ {\bf k}^2 + m^2 a(\tau)^2}} \; e^{i{\bf k}({\bf x}-{\bf x}')}  \;    d{\bf k} 
\right]
=
\frac{m^2 a^2}{(4\pi)^2}   \at -\log {\frac{(m a)^2}{\la^2}}  +  c  \ct 
$$
which has to be understood in the sense of distributions, while $c$ is a constant.
Hence, as a side remark, it is clear that the Hadamard regularization  and the first order adiabatic regularization \cite{Parker} for the computation of the expectation values of 
$\phi^2$ coincides up to the choice of a renormalization constant.
Furthermore, with a judicious choice of $\la$ (proportional to $m$), a direct computation shows that for every $x$ and $x'$ lying on the same Cauchy surface $\Si_\tau$ 
 $$
\lim_{x'\to x} \aq\cH^0_\bM(x,x')-  \frac{1}{(2\pi)^3} \int  \Theta(|{\bf k}|-m a(\tau) ) \at \frac{1}{2|{\bf k}|} - \frac{m^2 a(\tau)^2}{4 |{\bf k}|^3 }  \ct 
 \; e^{i{\bf k}({\bf x}-{\bf x}')}  \;    d{\bf k}
 \cq
 =  \frac{m^2 a^2}{8\pi^2} ( -\log{a} + c')\;,
 $$ 
and hence we could use the right hand side to regularize the states in order to obtain the expectation value of $\phi^2$.
Once again we incorporate the constant  $c'$ in the renormalization constant $\beta$, that we shall fix a posteriori according to the discussion given above.

Eventually, also in the computation of the expectation value of $\phi^2$, we are free to restrict both $x$ and $x'$ to the same Cauchy surface, determined by $\tau_1=\tau_2=\tau$ and, after the subtraction, we can perform the coinciding point limit
\begin{gather}
\langle\phi^2(x')\rangle_{\bom_{1,0}}:= 
\nonumber
\\
\lim_{x\to x'}  \frac{1}{(2\pi)^3a(\tau)^2}\int d{\bf k}\;  e^{i{\bf k}(x-x')} 
\aq\overline{\chi}_k(\tau)\chi_k(\tau) - 
\Theta(|{\bf k}|-m a(\tau) ) \at \frac{1}{2|{\bf k}|} - \frac{m^2 a(\tau)^2}{4 |{\bf k}|^3 }  \ct
\cq
+\beta m^2,
\label{Phi2}
\end{gather}
where $\beta$ can now be fixed to the value $-1/(8\pi^2)$ by the requirement that $\phi^2$ vanishes in the limit of flat spacetime, furthermore, the integrals present above are to be intended in the distributional sense.
Notice that the $\log(a)$ terms are not manifestly present anymore in the expectation value of $\phi^2$. 
We conclude this section with a remark. Since the map $t\to \tau$ is a diffeomorphism, the previous expression can be easily given in cosmological coordinates, just employing $a(t)$ and $\chi_k(t)$ at the place of $a(\tau)$ and $\chi_k(\tau)$, where, as discussed above,
$a(\tau)=(a \circ \tau^{-1})(\tau) $ and $\chi_k(t):=(\chi_k\circ \tau) (t)$.

\subsection{Recursive constructions and bounds satisfied by $\chi_k$ and by the auxiliary functional $\Phi$}\label{chiphi}

The aim of the present section is to find suitable bounds satisfied by $\bom_{1,0}(\phi^2)$, for small values of $a$ and close to $t=0$.
In order to accomplish this task we need some better control on the explicit form of the $\chi_k$ appearing in \eqref{Phi2}.
Hence, here we shall give a recursive construction for $\chi_k$, used in the definition of $\om_{1,0}$
using $m$ as a perturbation parameter.
While the analysis of the state, and of the modes $\chi_k$ in particular, is easier if performed in the conformal time $\tau$, the form of the differential equation \eqref{semiclassical} suggests the use of the cosmological time $t$.
Here, as previously said since the map $\tau$ from the cosmological time to the conformal is a diffeomorphism, without ambiguity we shall write $\chi_k(t)$ for $(\chi_k\circ \tau) (t)$ where $\chi_k(\tau)$ is a solution of \eqref{eqmotoT} with the initial conditions \eqref{initialconditionchi}. 
Following similar ideas to the one presented in \cite{FH} though in another context, we can write $\chi_k(\tau)$ by means of perturbation theory, hence in term of a Dyson series over the massless solution $\chi_k^0$ and treating $m^2a(\tau)^2$ as the perturbation potential
(a similar construction can be found in the proof of Theorem 4.5 in \cite{DMP2}).
Finally, since we would like to use the cosmological time instead of the conformal one, we can simply compose $\chi_k$ with $\tau$ (writing $\chi_k(\tau(t))$) and then change the coordinates in every integral appearing in that series from the conformal time to the cosmological one.
%
%
Following this procedure we obtain
\beq\label{seriechi}
\chi_k[a,\tau](t)=\sum_{n=0}^\infty \chi^n_k[a,\tau](t)\;,
\eeq
where,
\beq\label{step}
\chi_k^n[a,\tau](t)=  -\int_{0}^t \frac{\sin(k(\tau(t)-\tau(t')))}{k}   a(t') m^2 \chi_k^{n-1}[a,\tau](t')\; dt' \;,
\eeq
and
$$
\chi_k^0[a,\tau](t)=  \frac{e^{-ik\tau(t)}}{\sqrt{2k}} \;.
$$
The advantage of using the cosmological time is in the fact that
it becomes clear how  $\chi_k$ depends on $\tau$.
Notice that $\tau$ appears only in  $\sin(k(\tau(t)-\tau(t')))$  of \eqref{step} and in the form of $\chi_k^0$.
This decrease the number of functional derivatives that needs to be performed in the computation of the functional derivatives of $\chi_k$ with respect to the small variations $\de X$ of $X$.
%
The first thing we would like to check is that the perturbative construction is well defined, namely let us show that the sum in \eqref{seriechi} converges uniformly.
We shall state the result about convergence in the following proposition and,
although the proof is almost straightforward we would like to give it since it contains some useful remarks that we shall use also later.

\begin{proposizione}\label{chiconvergence}
Let $(M,g)$ be the flat FRW spacetime determined by $X\in\cB_c$ according to the statement of proposition \ref{functorialspacetime}, where $a$ is its scaling factor and $\tau$ the corresponding conformal time.
%
Then the series \eqref{seriechi}, constructed over $(M,g)$, converges absolutely to  ${\chi_k}$
on $[0,t_0]$, where $\chi_k$ is a solution of equation \eqref{eqmotoT}, when read in conformal time, and, moreover, it enjoys the initial condition \eqref{initialconditionchi}.
Furthermore $\chi_k^n$, given in \eqref{seriechi}, and 
$\chi_k$ satisfy the following bounds
$$
|\chi_k^n(t)|\leq \frac{1}{\sqrt{2k}}\frac{(mt)^{2n}}{n!} \;,
\qquad
|\chi_k(t)| \leq \frac{1}{\sqrt{2k}}
\exp
{\at \frac{m^2 a(t)t}{k} \ct}  
 \;, 
\qquad
|\chi_k(t)| \leq 
\frac{1}{\sqrt{2k}}\exp{ \at m^2 t^2 \ct }  \;.
$$ 
Introducing the following notation 
$$
\chi^{>n}_k[a,\tau](t)=\sum_{i=n+1}^\infty \chi^i_k[a,\tau](t)\;,
$$
we have also that 
$$
\left| \chi^{>n}_k (t) \right| \leq {C_n(t)}\frac{1}{k^{n+\frac{1}{2}}} \;. 
$$
\end{proposizione}
\begin{proof}
By direct inspection, under the hypotheses stated in the proposition, if the series converges, it has to tend to a solution of the equation of motion \eqref{eqmotoT} which satisfies the wanted initial conditions \eqref{initialconditionchi}.
More precisely, notice that 
$$
\at\frac{\pa^2}{\pa \tau^2} +k^2\ct \chi_k^n(\tau)+m^2a^2 \chi_k^{n-1}(\tau)=0\; \qquad \text{and} \qquad \at\frac{\pa^2}{\pa \tau^2} +k^2\ct \chi_k^0(\tau)=0\;.
$$
Hence, let us apply the operator realizing the equation of motion \eqref{eqmotoT} to the truncated series ($\chi_k-\chi_k^{>n}$). Using the inequalities given in the proposition, we obtain that the absolute value of the reminder of the latter operation is smaller then $a_0^2m^2 (mt_0)^{2n} / (n!\sqrt{2k})$, which tends uniformly to zero over $[0,t_0]$ for large $n$.
At the same time, for every $n>0$, $\lim_{\tau\to-\infty}e^{ik\tau} \chi_k^{n}(\tau)=0$  and the same holds for the first $\tau-$derivative of $\chi_k^{n}$.

What remains to be shown is the convergence and the validity of the estimates given in the statement of the proposition.
We proceed analyzing the $n-$th element of the sum. Looking at the recursion relations \eqref{step}, it is a straightforward task to obtain 
$$
| \chi_k^n(t) | \leq   
\int_{0}^t  \frac{a(t')}{k} \;  m^2 | \chi_k^{n-1}(t') |\; dt' \;,\qquad
| \chi_k^n (t)| \leq   
\int_{0}^t  (\tau(t)-\tau(t'))  a(t') m^2 | \chi_k^{n-1}(t') |\; dt' \;.
$$
The exponential bounds presented in the proposition can be obtained noticing that, $W(t_1,\dots,t_n)$ being any totally symmetric integrable function,
\begin{gather}\label{partition}
\int_{0 \leq t_1\leq t_2 \dots \leq t_n \leq t } \!\!\!\!\! 
W(t_1,\dots,t_n) 
\; dt_1\dots dt_n 
= 
\frac{1}{n!}  \int_{[0,t]^n} \!\!\!\!\!
W(t_1,\dots,t_n)\;  
dt_1\dots dt_n \;.
\end{gather}
Hence, out of \eqref{partition},
for any $V(t)$,
\begin{gather*}
1
+
\int^t_{0}\!\!\! dt_1 V(t_1)  
+ 
\int^t_{0}\!\!\!dt_1  V(t_1) 
\int^{t_1}_{0}\!\!\!dt_2  V(t_2)
+ \dots
=
\exp \int^t_{0} dt_1 V(t_1).
\end{gather*}
Together  with 
$| \chi^0_k | ={(2k)^{-\frac{1}{2}}}$,
the last observation entails both the absolute convergence of the series and the following bounds 
$$
|\chi^n_k(t)| \leq \frac{1}{\sqrt{2k} \; n!}
\at  \int^t_{0}  {m^2} (\tau(t)-\tau(t')) a(t')   dt' \ct^n\;,
$$
$$
|\chi_k(t)| \leq \frac{1}{\sqrt{2k}}
\exp \int^t_{0}  \frac{m^2}{k} a(t')   dt'\;, 
\qquad
|\chi_k(t)| \leq \frac{1}{\sqrt{2k}}
\exp \int^t_{0}  {m^2} (\tau(t)-\tau(t')) a(t')   dt'\;. 
$$
Finally, since $a(t)$ is growing monotonically and since $\tau(t)-\tau(t') = \int^t_{t'}a(t'')^{-1} dt'' $,
the estimates presented in the 
proposition descend straightforwardly.

\end{proof}

The previous proposition yields suitable estimates for $\chi_k[a,\tau](t)$. We shall present the derivation of further useful estimates also for its functional derivative $D\chi_k[a,\tau,\DX](t)$ with respect to the small variation $\DX$ of $X$, in appendix \ref{stimechi}.
%
%
We shall conclude this subsection, by discussing some properties of the auxiliary functional $\Phi$ on $\cB_c$ defined as   
\beq\label{Phi}
\Phi[X] := 2\pi^2 a^2  \langle  \phi^2 \rangle_{\bom_{1,0}}  
\eeq
which corresponds to the expression \eqref{Phi2} multiplied by $2\pi^2 a^2$ and evaluated in the proper time $t$ instead of the conformal time $\tau$.
For that expression we have the following proposition whose long and technical proof can be found in appendix \ref{stimephi}. 

\begin{proposizione}\label{a2phi2bound}
The following inequality holds on $\cB_c$
$$
|\Phi[X](t) | \leq  C_1(a_0,c,t_0) t\;.
$$
Furthermore, the functional derivative\footnote{A nice introduction of these concepts can be found in \cite{Hamilton}.} of $\Phi$, with respect to the small perturbation $\DX$ of $X$, satisfies the following inequality
$$
\left|D\Phi[X,\DX](t)\right| 
\leq C_2(a_0,c,t_0) t_0  \|\DX\|_\mB\;,
$$
where $C_1(\ell t_0,c,t_0)$ and $C_2(\ell t_0,c,t_0)$ are uniformly bounded for $t_0\to 0$ when $\ell$ and $c$ are fixed. Above $\|\cdot\|_\mB$ is the norm of the Banach space $\mB$ introduced in \ref{Bc}.
\end{proposizione}

\subsection{Existence of exact solutions of the semiclassical Einstein equations out of initial data given at the beginning of the universe}

In this subsection, we shall present the main result of this paper, namely the existence of solutions of the semiclassical Einstein equations uniquely determined by some initial conditions and by the form of the state $\bom_{1,0}$ in particular. Later on we shall generalize the result to other Hadamard states in $\mC(M)$.
The proof we shall give is very similar to the Picard-Lindel\"of proof of existence of solutions of first order differential equations together with the complication that the potential is a functional of the solution and not simply a function. Nonetheless, we shall show that it will be possible to use similar methods thanks to the estimates provided by proposition \ref{a2phi2bound}.
Hence, let us start stating the following proposition which we shall use in order to prove that there exist solutions of \eqref{semiclassical},

\begin{proposizione}\label{isacontraction}
Fix $\ell$ and let $a_0=\ell t_0$, then, if $t_0$ is sufficiently small, the image of $\cB_c$ under the map $\cT$ defined in \eqref{contraction}, is contained in $\cB_c$. Furthermore, on $\cB_c$, $\cT$ is a contraction.
\end{proposizione}
\begin{proof}
First of all we have to show that 
$\cT$ maps $\cB_c$ to $\cB_c$. In this respect, proposition \ref{HadamardinBc} and the subsequent remark ensure that we can compute 
$\Phi$ for every element $X$  of $\cB_c$; hence we have to prove that 
$$
\left\| \cT\at X\ct - f_0      \right\|_\mB \leq  \frac{1-c}{2}\;,
$$
where the norm $\|\cdot\|_\mB$ is the one introduced in definition \ref{Bc} of $\cB_c$ and $f_0$ is the center of the ball $\cB_c$ introduced in \eqref{center}. 
Fixing the two constants $C_1 = 120 m^2 H_c^{-2}$ and $C_2= (1-c) H_c^{-2} $, the previous inequality is equivalent to  
$$
-C_2  +(2-c){X^2}(t) \leq  
 C_1 \frac{X^4(t)}{a^2(t)} \Phi(t) \leq { X^2(t)}\;\qquad \forall t\in[0,t_0].
$$
Since ${X^2} \leq  t^2 \leq t_0^2 $, if $t_0$ is sufficiently small, 
this chain of inequalities is satisfied if
$$
\left|  \Phi (t)  \right| \leq \frac{1}{C_1} \left| \frac{a^2(t)}{X^2(t)} \right|
  \; \qquad \forall t\in[0,t_0]
$$
which holds true for a certain $t_0$ thanks both to the bounds, satisfied by the elements of $\cB_c$, given in proposition \ref{someinequalities} and to the  estimates for $|\Phi[X](t)|$ found in proposition \ref{a2phi2bound}.

The next step consist to show that $\cT$ is a contraction on $\cB_c$  with respect to the norm of $\mB$, that is, for every $X_1$ and $X_2$ in $\cB_c$
$$
\left\| \cT\at X_1\ct - \cT\at X_2\ct   \right\|_\mB 
< C \left\| X_1 - X_2   \right\|_\mB\;,
$$
with $0<C<1$.
This property descends once more from the results stated in proposition \ref{a2phi2bound}, noticing that 
for every $0\leq\la\leq 1$ we can consider 
$$
{X_\la}:= \la X_1   + (1-\la) X_2  = 
{X_2}
+
 \la \aq X_1 - X_2 \cq   \;,
$$
where ${X_\la}$ is in $\cB_c$. Furthermore, indicating by $\DX$ the difference $X_1-X_2$, we have
$$
\left|\frac{d}{dt} \cT\at {X_1}\ct(t) - \frac{d}{dt} \cT\at {X_2} \ct(t) \right|
\leq  
\left| \int_0^1    \frac{d}{d\la}    \frac{d}{dt} \cT\at {X_\la}\ct(t)   d\la \right |  
\leq
\sup_{\la\in [0,1]} 
\left|  D \frac{d}{dt} \cT\at {X_\la}, \DX \ct(t)   \right|\;.
$$
In order to conclude the proof we have to analyze the functional derivative of $\frac{d \cT}{dt}$ in $\cB_c$. In this latter step, the contribution which requires some care is the one arising from the functional derivative of $X^4 a^{-2}  \Phi$, which, using both the estimates \eqref{Da}  derived in the the appendix and the result of proposition \ref{a2phi2bound}, can be shown to satisfy the inequality 
$$
\sup_{t\in[0,t_0]}\left| D\at \frac{X^4 \; \Phi}{a^2}\ct \aq X,\DX\cq (t)\right| \leq C_3 t_0 \| \DX  \|_\mB \;,
$$
for every $X$ in $\cB_c$, where $C_3$ is some constant. It is not difficult to show that similar estimates hold also for the other contributions to the functional derivative of $\frac{d{\cT}}{dt}$. 
Hence, there is certainly a sufficiently small $t_0$ for which $C<1$ and hence the map $\cT$ acts as a contraction therein.
\end{proof}

\begin{teorema}\label{esistenza}
Fix $\ell$ and let $a_0=\ell t_0$. If $t_0$ is chosen sufficiently small, it exists a unique solution $X= H^{-1}$ of \eqref{semiclassical} in $\cB_c$ which can be constructed recursively starting from the massless solution $X_0:=H_0^{-1}$ implicitly given in \eqref{massless-solution}. In other words the sequence
$$
X_n := \cT\at X_{n-1} \ct \;,
$$
converges in $\cB_c$ to a solution of the semiclassical Einstein equations.
\end{teorema}
\begin{proof}
On account of proposition \ref{isacontraction},
the proof descends out of a straightforward application of the Banach fixed point theorem to the contraction $\cT$ on $\cB_c$.
\end{proof}

Before concluding this subsection we would like to briefly comment on the result obtained above.
First of all notice that the method to obtain the solution is constructive, in the sense that for every $n$, $X_n$ is closer to the solution then $X_{n-1}$.
Furthermore, notice that if we start from $H_0$, which is a smooth function in $\cB_c$, every $H_n$ is also smooth, because $\cT$ maps smooth functions to smooth ones. 
Unfortunately out of the previous theorem, we cannot conclude that the found solution is smooth. In order to address this problem we should discuss the convergence of the series $X_n$ in the Fr\'echet spaces of smooth functions $C^\infty[0,t_0]$, but this is out of the scope of the present paper.
For our purpose, on account of the preceding discussion, it is enough to notice that we can find a smooth spacetime $X_n$ as close as we want, in the norm of $\mB$, to the the exact solution $X$ provided by the previous theorem.
Furthermore, the regularity displayed by the generic element of $\cB_c$ is enough in order for equation \eqref{semiclassical} to be meaningful.

As discussed at the beginning of this section we have used only part of the information present in the semiclassical Einstein equations, namely only the trace. What is left is completely set once the parameter $a_0$ is fixed in agreement with the expectation value of $T_{00}$.
Let us start noticing that in equation \eqref{semiclassical} nothing depends explicitly on $a_0$ and actually, also in 
$\langle \phi^2 \rangle_{\bom_{1,0}}$ as given in equation \eqref{Phi2}, $a_0$ does not really appear. 
The latter statement becomes evident, re-scaling everything with respect to $a_0$, that is, using $\overline{k}:= a_0^{-1} k$, $\overline{\tau}:=a_0 \tau$,  $\overline{a}:= a_0^{-1} a$ and 
$\chi^r_{\overline k}(\overline{\tau}) = \sqrt{a_0} \chi_{k}({\tau}) $
which now satisfy the equation
$$
\frac{\pa^2}{\pa \overline{\tau}^2} \chi^r_{\overline k}(\overline{\tau}) + \at \overline{k}^2 + m^2\overline{a}\ct \chi^r_{\overline k}(\overline{\tau}) =0 \;.
$$
Hence, once a solution of \eqref{semiclassical} is obtained, we
can always find other solutions by choosing a different $a_0$.
In this sense, $a_0$ is really a free parameter that can be 
fixed using the information stored in the expectation value of  $T_{00}$.
Actually, it can be shown that $\bom_{1,0}(T_{00})$ diverges when $H=X^{-1}\to \infty$; hence, the choice the state $\bom_{1,0}$ really provides the exact solution presented here, in the sense that the latter condition permits really to start the iterative construction of $H=X^{-1}$ out of the upper brunch $H_0$ of the massless solution \eqref{massless-solution}.

We conclude the present section noticing that the same result about the uniqueness and the existence of the solution of the semiclassical Einstein equations holds also for other choices of Hadamard states in $\mC(M)$ (see definition \ref{statiomogenei}) provided suitable conditions are satisfied. 
We have in fact the following theorem

\begin{teorema}\label{esitenzaaltristati}
Let $\bom_{A,B}\in\mC(M)$ be constructed out of $A$ and $B$ 
in $C^2(\bR^+)$,
%
and such that, with respect to the auxiliary function $f(k):=\overline{A(k)}B(k) k^2$, the following conditions are satisfied 
\begin{itemize}
\item[a)] $B$ is rapidly decreasing,
\item[b)] $\overline{A(k)}B(k)$ and $|B(k)|^2$ are contained in $L^1(\bR^+,kdk)$,
\item[c)]  $f(0)=f'(0)=0$,
\item[d)] $f'',f'\in L^1(\bR^+,dk)$.
\end{itemize}
Let $a_0=\ell t_0$; then, if $t_0$ is sufficiently small,  a unique solution of the semiclassical Einstein equations exists.
\end{teorema}
\begin{proof}
The proof of the present theorem can be performed following the one of theorem \ref{esistenza}. For this reason, here, we shall discuss only the differences. 
Let us start noticing that what changes in this case is the value of $\langle\phi^2\rangle_{}$.
More precisely, instead of dealing with the map $\cT$, we have to deal with $\cT'$
%
which is defined as follows
$$
\cT'\at X\ct:=\cT (X) + \frac{X^4}{1-H_c^2\; X^2}\frac{\Phi_{A,B}}{a^2}
$$
where 
$
{\Phi_{A,B}}:= 2\pi^2 a^2 \at \langle \phi^2 \rangle_{\bom_{A,B}} -\langle \phi^2 \rangle_{\bom_{1,0}} \ct
$ 
which can be rewritten as 
$$
\Phi_{A,B}(t)=    
 \int_0^\infty \left[
2 
|B(k)|^2\;\overline{ \chi_k}  \chi_k(t)\; 
+
\overline{A(k)} {B(k)} \;\overline{\chi_k}\overline{\chi_k}(t) + 
{A(k)} \overline {B(k)}\;\chi_k\chi_k(t)\; \cq 
\; k^2\; dk \;.
$$
Since we already have the estimates for $\chi_k-\chi^0_k$ given in the appendix \ref{stimechi}, it is useful to divide $\Phi_{A,B}$ in two parts, namely
$$
\Phi_{A,B} = \Phi^0_{A,B}+\Phi^+_{A,B} 
$$
where $\Phi^0_{A,B}$ is constructed as $\Phi_{A,B}$ with $\chi_k^0$ in the place of $\chi_k$.
Let us introduce the closed set $\cB'_d$ formed by the elements of $\mB$ such that 
$$
\cB'_d:=\left\{ X\in \mB \; ,\;   1 - t d\leq \frac{d}{dt}{X}(t)  \leq 1 + t d
\right\} \;.
$$
We have that $\Phi^+_{A,B}$  satisfies on $\cB'_d$, similar inequalities as the ones stated in proposition \ref{a2phi2bound} for $\Phi$.
In order to verify the last statement, we could make profitable use of the fact that $B(k)$ is rapidly decreasing and that $B(k)\chi_k$ and $A(k)\chi_k$ are locally square integrable with respect to the measure $k^2dk$. Hence
the first bound 
$$
|\Phi^+_{A,B}(t)|\leq C t  \;,\qquad \forall t\in [0,t_0]
$$
descends straightforwardly from the inequalities in 
proposition \ref{chiconvergence}, while the second one, about the variation of $\Phi^+$ under the small perturbation $\DX$ of $X$, 
$$
|D\Phi^+_{A,B}(t)|\leq C t_0 \| \DX\|_\mB,\qquad \forall t\in [0,t_0]
$$
can be proved using the results stated in the appendix \ref{stimechi}.
$\Phi^0_{A,B}$ does not satisfy similar inequalities, and it is for such a reason that we have introduced $\cB'_d$, a closed set different then $\cB_c$.
For $\Phi^0_{A,B}$, there exists a constant $\de_1$ such that   
$$
|\Phi^0_{A,B}(t)|\leq \de_1 ,\qquad \forall t\in [0,t_0]
$$
and this inequality descends straightforwardly form the properties a) and b) stated in the hypotheses and from the fact that $|\chi_k^0| k $ is constant.
The variation of $\Phi^0_{A,B}$ under the small perturbation $\DX$ of $X$ involves only the variation of $e^{2ik\tau}$, present in the second and third term of $\Phi^0_{A,B}$ .
From properties c) and d), it descends the following uniform decay 
$$
\left| \int \overline{A(k)}B(k) k^2 e^{2ik\tau(t)} dk
\right| \leq \frac{C}{\tau^2(t)} \; ,\qquad \forall t\in [0,t_0].
$$ 
Furthermore, operating in a similar way as in appendix \ref{A1}, since $|\DX(t)|\leq t \|\DX\|_\mB $, we obtain the uniform estimate
$$
\sup_{t\in[0,t_0]}\left|\frac{D\tau(t)}{\tau^2(t)} \right| \leq C 
\sup_{t'\in[0,t_0]}\left| \frac{t'^2}{X^2(t')}\right| \|\DX\|_\mB
$$
valid for sufficiently small $t_0$ in $\cB'_d$ and for some constant $C$.
Combining these arguments we easily obtain
$$
|D\Phi^0_{A,B}(t)|\leq C \| \DX\|_\mB.\qquad \forall t\in [0,t_0]
$$
Finally, let us notice that the results given in proposition \ref{a2phi2bound} hold also on $\cB'_d$.

Although the behavior of $\Phi^0_{A,B}$ is worse then that of 
$\Phi^+_{A,B}$ or of $\Phi_{A,B}$, the inequalities given above 
are sufficient in order to obtain the existence and uniqueness 
of the solution in $\cB'_d$.
This is possible because, on $\cB'_d$,
$ |X a^{-1}(t)| \leq C $ which is not valid for the generic 
element of $\cB_c$.
Actually, for a sufficiently large $d$ (depending on $\de_1$ in particular) and a sufficiently small $t_0$,
the map $\cT'$ introduced above, is contractive in $\cB'_d$ as it can be seen adapting 
the proof of proposition \ref{isacontraction} to encompass the case under the present investigation.
Out of this observation the proof of the present proposition can then be easily concluded, by applying the Banach fixed point theorem on $\cB'_d$. 
\end{proof}

\se{Further considerations on the semiclassical solutions}

\subsection{Expectation value of $\langle \phi^2 \rangle$ at late times}

Out of direct inspection we notice that, at late times, 
the massless solution \eqref{massless-solution} tends to the de Sitter spacetime. 
We would like to see if such behavior persists also for massive fields\footnote{See also the work \cite{Nadal}.}.
To this end notice that in \cite{DFP} it is argued that, in the case of a massive conformally coupled scalar field, if the expectation value of
$\phi^2$ has the form $\beta m^2 +\alpha R$ then, the same behavior at late times can be inferred.
Such an estimate was obtained analyzing the adiabatic approximation for the state.
It would be desirable to see if the assumption made in
\cite{DFP} for the expectation value of $\phi^2$ is reliable for the exact solution of the semiclassical Einstein equations.

We start noticing that, in the case of an exact de Sitter spacetime, the expectation values of $\phi^2$ in the Bunch Davies \cite{BD} state has exactly that behavior, and hence, 
there is only one choice of the cosmological constant which solves the semiclassical Einstein equations.
Unfortunately, it is very difficult to compute the evolution of the expectation value of $\phi^2$ in the exact solution we have found before.
Despite the difficulties present in the analysis of the semiclassical solution at finite times, we would like to stress that 
if a de Sitter phase exists at late times, it has to be exactly the one found in \cite{DFP}.
Such a remark originates from the fact that the expectation value of 
 $\phi^2$, in an expanding flat universe ($H>0 $), is state independent (at least on the class of homogeneous pure Hadamard states) as asserted in the following theorem. 

\begin{teorema}
In an expanding $H> 0$ flat FRW with smooth scale factor $a$ which diverges in the limit $t\to\infty$, choosing a quasi-free state $\bom_{A,B}$ whose two-point function is as in \eqref{omegaAB} and of Hadamard form,
the expectation value of $\phi^2$ in $\bom_{A,B}$ equals the one computed on $\bom_{1,0}$.
\end{teorema}

\begin{proof}
The proof of this theorem can be obtained considering the form 
$\chi_k$. 
Since both $\bom_{A,B}$ and $\bom_{1,0}$ are Hadamard states, it holds for their two-point functions
\begin{gather}
\om_{A,B}(\phi^2(x)) - \om_{1,0}(\phi^2(x)) = 
\nonumber
\\
\label{differenzastati}
 \frac{1}{(2\pi)^3}\frac{1}{a(t)^2}   
 \int \left[
2 
|B(k)|^2\;\overline{ \chi_k}  \chi_k(t)\; 
+
\overline{A(k)} {B(k)} \;\overline{\chi_k}\overline{\chi_k}(t) + 
{A(k)} \overline {B(k)}\;\chi_k\chi_k(t)\; \cq 
\; d{\bf k}
\end{gather}
where $x=(t,{\bf x})$ in cosmological coordinates and  $k=|{\bf k}|$. Because of theorem \ref{differenza}, $B(k)$ is rapidly decreasing and $A(k)$ of polynomial growth, hence we can pass the coinciding point limit under the sign of $k$-integration. 
We aim to prove that the preceding expression vanishes in the limit $a\to\infty$. To this end we shall show that the previous $k$ integration is bounded in time.
To prove the last statement we have to 
analyze the behavior of $\chi_k$ for large times. To this avail, let us  consider the functions $\chi_k$ in cosmological time ($\chi_k(t):=(\chi_k\circ\tau)(t)$), and let us notice that they have to satisfy the following equation
$$
\ddot\chi_k +H\dot\chi_k + \at \frac{k^2}{a^2}+m^2 \ct \chi_k =0
$$
where $\dot\chi_k$ and $\ddot\chi_k$ indicate respectively the first and second derivative of $\chi_k$ with respect to the cosmological time $t$. Hence the previous equation descends straightforwardly out of a change of variables in \eqref{eqmotoT}.
Let us introduce the positive quantity
$$
Q(t,k) := \overline{\dot\chi_k} \dot\chi_k(t) + \at \frac{k^2}{a^2}+m^2 \ct \overline{\chi_k}\chi_k(t) \;.
$$ 
We would like to show that, at fixed $k$, $Q(t,k)$ decreases in time. To this end notice that, since $H\geq 0$ 
$$
\dot{Q}(t,k) = - 2H | \dot\chi_k(t)|^2   -2 \frac{k^2}{a(t)^2} H(t) |\chi_k(t)|^2 \leq 0,
$$
where we have indicated by $\dot{Q}(t,k)$ the $\frac{\pa}{\pa t}{Q}(t,k)$.
Hence, for every $t>t_0$, we have that  
$$
Q(t,k) \;=\; Q(t_0,k) + \int_{t_0}^{t} \dot Q(t',k) dt'\;  \leq\;\; Q(t_0,k)\;. 
$$
Out of the form of $Q(t,k)$ we obtain that, for every $t>t_0$, 
$$
| \chi_k(t) |^2  \leq  \frac{Q(t_0,k)}{m^2}\;. 
$$
It is now possible to use the preceding inequality to estimate the ${\bf k}$-integral present in \eqref{differenzastati},
and since both $Q(t_0,k) |B(k)|^2$ and $Q(t_0,k) |A(k) B(k)|$ are  integrable in $k$, we have that it exists a positive constant $C(t_0)$,  such that, for every $x$ in the future of the surface $\Si_{t_0}$,
$$
\left|
\om_{A,B}(\phi^2(x)) - \om_{1,0}(\phi^2(x)) 
\right|
\leq \frac{C(t_0)}{a(t)^2} \;.  
$$
Out of the preceding remark, we can 
straightforwardly evaluate the limit $t\to\infty$ and, since 
$a(t)$ diverges per hypothesis therein, the difference vanishes.
\end{proof}

\subsection{Variance of $\langle T\rangle $ when smeared on  large volume regions}

As discussed in the introduction, the semiclassical Einstein equations  are valid when the quantum fluctuations of the expectation value of the stress tensor are negligible.

In this paper, in order to obtain the dynamics of the spacetime, due to the large symmetry present in the model under investigation, we have used only the trace of the stress tensor.
Notice that the anomalous terms in the trace  are $c-$numbers; hence, they have vanishing fluctuations, and this already means that, in the case of massless fields, the found semiclassical equation and its solution are meaningful at every times (not only when the curvature is small).

Furthermore, in the case of a massive field conformally coupled with the curvature, the source for the fluctuations in the trace of the stress tensor can only be the expectation value of $\phi^2(x)$.
Unfortunately we have to notice that point-like fields like $\phi^2(x)$, although represented by smooth functions, could have divergent variance.
This is indeed the case even if the Hadamard regularization is employed, as can be seen noticing that in  
\beq\label{variance}
\phi^2(x)\star_\cH\phi^2(y) = \phi^2(x) \phi^2(y) +4  \cH(x,y)\phi(x)\phi(y)+ 
2 \cH(x,y) \cH(x,y)\;
\eeq
the product $\cH(x,y) \cH(x,y)$ diverges in the coinciding point limit.
Above $\cH$ is the Hadamard singularity and $\star_\cH$ is the product in the extended algebra of observables $\mF(M)$, as discussed in \cite{BF,BDF}. Some further comments on such a  product are given in appendix \ref{deformationquantization}. We shall use that expression in the computation of the variance of $\phi^2(f)$, namely on a stat $\bom$ it looks like
\beq\label{variance}
\De_{\bom}(\phi^2(f)):=\bom(\phi^2(f)\star_\cH\phi^2(f)) - \bom(\phi^2(f)) \bom(\phi^2(f)) \;.
\eeq
Furthermore, if the state is quasi-free, indicating by $W$ the smooth part of the two-point function, it becomes
$$
\int \aq 2 W(x,y) W(x,y) + 4 \cH(x,y) W(x,y) + 2 \cH(x,y) \cH(x,y) \cq
f(x) f(y) d\mu(x) d\mu(y)
$$
which, on the state $\bom_{1,0}$ we have chosen,  is equal to 
$$
2 \int \om^2_{1,0}(x,y) f(x) f(y) d\mu(x) d\mu(y).
$$
where $\om_{1,0}^2$ is the product of distribution $\om_{1,0}\cdot \om_{1,0}$.
Although $\om_{1,0}^2$ is a well defined distribution, due to the form of its wave front set,  its coinciding point limit is not well defined and actually it diverges.
Hence, the expectation value of $\phi^2(x)$ has divergent fluctuations, but, luckily enough, we can overcome this difficulty by a suitable smearing procedure we are going to introduce.

To this end, we could use once again the spatial symmetry and
consider a smearing function $f_{n_1,n_2}$ such that, for large $n_1$ and $n_2$ it tends to have support on the whole Cauchy surface $\Si_{\tau}$, and it is normalized in such a way that its integral on $M$ is equal to one.
More precisely, we consider 
a compactly supported smooth function $f\in C^\infty_0(M)$ centered in  $x_\tau:=(\tau,0)$, and such that 
$$
f(x_\tau)=1\;,\qquad \int_M f d\mu(g) = 1 \;,\qquad  0\leq f(x) \leq 1 \;.
$$
Notice that, in writing the preceding expression, we are using the conformal coordinates on $M$.
We can then generate the desired $f_{n_1,n_2}$ as
\beq\label{testf}
f_{n_1,n_2}(\tau',{\bf x}) =  \frac{n_1}{n_2^3} f\at n_1 (\tau'-\tau) +\tau , \frac{{\bf x}}{n_2}\ct \frac{\sqrt{|g(n_1(\tau'-\tau)+\tau,{\bf x}/n_2)|}}{\sqrt{|g(\tau',{\bf x})|}}
\eeq
and, as anticipated before, for large $n_1$,  $f_{n_1,n_2}$ tends to have support on $\Si_{\tau}$ while, for large $n_2$, even if its spatial support becomes larger and larger, $\int_M f_{n_1,n_2} d\mu(g) = 1$.

We can then smear the fields in \eqref{semiclassical} and subsequently in \eqref{variance} with $f_{n_1,n_2}$ and analyze the behavior of the smeared quantity for large $n_1$ and $n_2$. 
In other words,
notice that, since the trace of the Einstein tensor $G$ is smooth and equal to $-R$, in the limit $(n_1,n_2)\to \infty$  
$$
\lim_{(n_1,n_2)\to \infty}\langle g^{\mu\nu} G_{\mu\nu}, f^{\tau}_{n_1,n_2} \rangle  = -R(\tau)\;,
$$
and, thanks to the continuity properties of $R$, the result does not depend on the order in which the limits are taken.
The analysis of the trace of $\langle T\rangle$ follows similarly. In fact, the anomalous part of the trace depends continuously on the coefficients of the metric, and its derivatives, up to the second order, while the expectation value of $\langle\phi^2(x)\rangle $ has vanishing wave front set and, hence, it is continuous too.
Thus, the equation $-R=8\pi \langle T \rangle_{\bom_{1,0}}$, together with its solutions, holds exactly in the same way for the smeared quantity.

We proceed to analyze the variance of these quantities, and we notice immediately that the variances of the geometric entities present in \eqref{semiclassical} vanish for large $n_1$ and $n_2$, no matter in which order the limits are performed. 
We have to be more careful when analyzing the variance of the expectation values of $\langle\phi^2\rangle$ but, luckily enough, we have the following theorem which shows that the fluctuations of $\phi^2$ vanish if the limits are taken in a suitable order.

\begin{teorema}
Let $f_{n_1,n_2}$ be some smearing functions constructed as in \eqref{testf} out of some $f$ with the properties stated above. Fix $n_1$, then the variance of $\phi^2(f_{n_1,n_2})$, computed according to  formula
\eqref{variance} and evaluated on the state $\bom_{1,0}$, vanishes for large $n_2$ in the weak sense, that is 
$$
\lim_{n_2\to\infty }\De_{\bom_{1,0}}(\phi^2(f_{n_1,n_2})) =0\;.
$$
\end{teorema}

\begin{proof}
First of all notice that, $\om_{1,0}$ is of Hadamard form and hence its wave front set satisfies the microlocal spectrum condition.
This implies that the pointwise product of $\om_{1,0}$ with itself, that we shall indicate by $\om^2_{1,0}$, is a well defined
 distribution because the H\"ormander criterion for multiplication of distributions is fulfilled (the sum of the wave front set of the distribution $\om_{1,0}$ with itself does not contain the zero section). Furthermore, theorem 8.2.10 of \cite{Hormander} yields that the wave front set of the product $\om^2_{1,0}$  has to be contained in
$(WF(\om_{1,0}) \cup \{ 0 \}) + (WF(\om_{1,0}) \cup \{ 0 \}) $, where the sum is meant in the cotangent space.
This implies, that the singular support of $\om^2_{1,0}$ can only contain points connected by a lightlike geodesic. Hence, we can divide in two parts the distribution 
$$
\om^2_{1,0} \at f_{n_1,n_2} \otimes f_{n_1,n_2} \ct :=
\om^2_{1,0}u\at f_{n_1,n_2} \otimes f_{n_1,n_2} \ct + \om^2_{1,0}u'\at f_{n_1,n_2} \otimes f_{n_1,n_2} \ct\;.
$$
Above we have split the distribution $\om^2_{1,0}$ multiplying the test functions by a suitable partition of unit $u+u'=1$ on $M\times M$
and using the linearity of $\om^2_{1,0}$.
%
We shall now discuss how we have to construct $u$ in order to have that the intersection of the singular support of $\om^2_{1,0}u'$ with the support of $f_{n_1,n_2} \otimes f_{n_1,n_2}$ is empty for every $n_2$ and for fixed $n_1$.  
Consider a compactly supported smooth function $\rho$ on $\bR^3$ equal to one on a neighborhood of the origin.
Then we shall indicate by $u$ the smooth function on $M\times M$, 
$$
u:((\tau,{\bf x});(\tau',{\bf x}')) \mapsto  \rho({\bf x}-{\bf x}'),
$$
where we have used standard coordinates introduced above.
Let us start analyzing  $\om_{1,0}^2u$, and 
suppose  to test it 
on a test function $h\in C^\infty_0(M\times M)$ of the following form
$$
h((\tau_1,{\bf x});(\tau_1,{\bf y})):=f_t(\tau_1,\tau_2)f_+({\bf x} - {\bf y})f_-({\bf x} + {\bf y})\;,
$$ 
where $f_t\in C^\infty_0 (I\times I)$ and $f_+,f_-\in C^\infty_0(\bR^3)$. 
Furthermore, these functions are chosen in such a way that 
$h$ is positive and, for a fixed $n_1$, $h\geq f_{n_1,1}\otimes f_{n_1,1} $.
Hence, due to the translation invariance satisfied by $\om^2_{1,0}$ and the multiplication by the cutoff $u$, the following continuity condition holds true
$$
| \om_{1,0}^2u\at 
h
\ct  | \leq  C \at  \sum_{|a|\leq q} \| D^a f_t\|_\infty   \ct \|f_+\|_1 \at  \sum_{|b|\leq q}\| D^b f_- \|_\infty \ct 
$$
for suitable values of $C$ and $q$.
Notice that, because of the localization introduced by $u$, $C$ does not depend on the support of $f_-$.
Out of this observation we have that, for $h_n\in C^\infty_0(M\times M)$ defined as $h_n((\tau_1,{\bf x});(\tau_1,{\bf y})):=h((\tau_1,{\bf x}/n);(\tau_1,{\bf y}/n))/n^6$,   
$$
\lim_{n\to \infty} \om_{1,0}^2u\at h_n \ct = 0.
$$
The latter limit implies also that  
$\lim_{n_2\to\infty}\om_{1,0}^2u(f_{n_1,n_2}\otimes f_{n_1,n_2})=0$, because, for fixed $n_1$, thanks to the special choice of 
$f_t$, $f_+$ and $f_-$ and to the positivity of $\om_{1,0}^2u$,
$$
\om_{1,0}^2u(f_{n_1,n_2}\otimes f_{n_1,n_2}) \leq \om_{1,0}^2u\at h_{n_2} \ct\;.
$$
 
We have now to analyze the second contribution to $\om_{1,0}^2$, namely $\om_{1,0}^2 u'$, and we notice that, per construction, the singular support of $\om_{1,0}^2 u' $ has vanishing intersection with the support of $f_{n_1,n_2} \otimes f_{n_1,n_2}$.
Hence, when tested on $f_{n_1,n_2} \otimes f_{n_1,n_2}$, $\om_{1,0}^2 u'$ has a smooth integral kernel, and, for every $n_2$ and for sufficiently large $n_1$,
$$
\om_{1,0}^2 u'(f_{n_1,n_2}\otimes f_{n_1,n_2})
$$ 
is represented by an ordinary integral. We can change the variable of spatial integration in such a way that the preceding operation can be  rewritten as
$$
\lim_{n_2\to\infty}\int (\om_{1,0}^2 u')(\tau_1,\tau_2,n_2 ({\bf x}-{\bf y})) n_1^2 f(n_1\tau,{\bf x})
f(n_1\tau',{\bf y})  d\mu(\tau_1,\tau_2) d{\bf x} d{\bf y}.
$$
Since $f$ has compact support and $(\om_{1,0}^2 u')$ is bounded, we can pass the limit under the sign of integration. We obtain that the limit for large $n_2$ vanishes provided that $\lim_{n_2\to\infty}(\om_{1,0}^2 u')(\tau_1,\tau_2,n_2 ({\bf x}-{\bf y}))=0$.

Due to the spatial translational and rotational invariance, it is enough to check this last requirement for $\om_{1,0}$ (without the square), for a single direction and when one of the points is in the spatial origin of the employed coordinate system. Hence, using the conformal coordinates,
let us consider $x:=(\tau_1,0)$ and $y:= (\tau_2,r,0,0)$, with $r>>0$. We can
rewrite the two-point function $\om_{1,0}$ as 
$$
\om_{1,0}(x,y) = \frac{2}{(2\pi)^2} \frac{1}{a(\tau_1)a(\tau_2)} \int_0^\infty   \frac{\sin(k r)}{r}\; \overline{\chi_k(\tau_1)}\chi_k(\tau_2)\; k\;d k\;,
$$
where, we have made use of the fact that $r$ is chosen to  be strictly positive and the fact that $(x,y)$ is not contained in the singular support of $\om_{1,0}$.
We can now split the $k-$integral on $(0,\infty)$ in two parts, as similarly done in the proof of proposition \ref{a2phi2bound}, namely  on the interval $(0,1)$ and on $(1,\infty)$.
Since  $\overline{\chi_k(\tau_1)}\chi_k(\tau_2)$ is locally integrable in the measure $k\; dk$, due to the estimates given in proposition \ref{chiconvergence} and, making use of the Riemann Lebesgue theorem, we can conclude that the contribution on $(0,1)$ vanishes when $r$ diverges.
Let us proceed to analyze the $k-$integral on the remaining interval $(1,\infty)$. Hence, we can now make use of the perturbative construction in $m^2$ for the $\chi_k$ given in \eqref{seriechi} and analyze the different contributions separately.
First of all we notice that the sum of the contributions to $\overline{\chi_k(\tau_1)}\chi_k(\tau_2)$ of order equal or larger to $m^4$ lie in $L^1(\bR,kdk)$, and, once more, due to results stated in proposition \ref{chiconvergence}, by the Riemann Lebesgue lemma, as $r$ diverges this contribution tends to zero.
We are left with the analysis of the order $1$ and $m^2$, 
The former consists of a well known distribution that vanishes for large $r$, whereas the latter (order $m^2$) can be estimated, using an argument similar to the one used to treat \eqref{cosxx}. Therefore, for large $r$ and for some constant $C$, this contribution can be shown to be bounded by $C r^{-1} (1+\log (r))$ which tends to zero when $r$ diverges.
\end{proof}

This discussion provides a new and stronger physical interpretation of the semiclassical Einstein equations on spacetime with large spatial symmetry as the cosmological one.

\se{Summary of the results and final comments}

In this paper we have rigorously constructed some solutions of the semiclassical Einstein equations in cosmological spacetimes. All these solutions display a phase of power law inflation after the initial singularity. 
We have also seen that, if a stable de Sitter phase is asymptotically  present in the future, then the acceleration of the latter does not depend on the particular homogenous pure state in which the quantum theory is evaluated, and hence it can be considered as a universal character.
Finally we have discussed the interpretation of these equations, showing that, when smeared on large spatial regions the equation  $-R= 8\pi \langle T\rangle$ continues to be valid, and,
in this case, since the fluctuations of $\langle T\rangle$ tend to vanish, it acquires an even stronger interpretation. 

We have been able to derive those results because we have
selected a class of spacetimes and we have used 
an universal quantization scheme for the matter fields on that class,
employing ideas typical of local covariance \cite{BFV}. 
Furthermore, here we have also been able to give unambiguously a quantum state on every element of the class of considered spacetimes, using and even generalizing, the results in \cite{DMP2,DMP4}.
Furthermore, the employed methods permit to characterize the quantum states in an intrinsic way.
In other words, using a lightlike initial surface and giving initial data thereon,
 we provided a prescription which can be used to construct states with a good ultraviolet behavior.

There are of course many questions that are still open. 
In principle, similar ideas could be used to treat also more complicated problems like, for example, the backreaction of  collapsing matter forming a black hole.
Another interesting aspect, that should be addressed in the future, is the relation between the results obtained here on the vanishing of the fluctuations by smearing the fields on infinite volume regions and the approach typical of the so called stochastic gravity \cite{HuVerd}.

\section*{Acknowledgments.} 
I would like to thank Romeo Brunetti, Claudio Dappiaggi,  Klaus Fredenhagen, Thomas-Paul Hack and Valter Moretti for useful discussions, suggestions and comments on the subject of this paper.
Supported in part by the German DFG Research Program SFB 676 and by the ERC Advanced Grant 227458 OACFT ``Operator Algebras and Conformal Field Theory".
 
\appendix



\section{Appendix}

\subsection{Deformation quantization and the extended algebra of fields}\label{deformationquantization}

The standard procedure for quantizing the system \eqref{KleinGordon} will be to choose the symplectic space of the theory $(\cS(M),\si)$ 
and then to construct the $C^*$-algebra generated by the Weyl operator, see for example \cite{Dimock}.
Here we shall follow another path, that is we introduce the quantization as it is custom in the so called deformation quantization \cite{BF,BDF}.
Let us define the 
off shell Borchers-Uhlmann algebra as 
$$
\mA(M) := \bigoplus_n^{\infty} C^\infty_0(M)^{\otimes_S n} 
$$
where only sequences with a finite number of elements are taken into account, $\otimes_S$ stands for the symmetric tensor product and the first space ($n=0$) in the previous direct sum is $\bC$.
The elements of $\mA(M)$ can be alternatively seen as the functionals on smooth field configurations $\varphi$.
To wit, to every $F:=\{F_n\}\in \mA(M)$, we can associate  
$$
F(\varphi):= \sum_n \frac{1}{n!} \langle F_n(x_1,\dots,x_n),\varphi(x_1)\cdot\dots \cdot\varphi(x_n)   \rangle\;,
$$
where $\varphi$ is in $C^\infty(M)$ and $\langle\cdot ,\cdot\rangle$ is the standard pairing. Out of this transformation the set $\mA(M)$ is interpreted as the set of functionals with smooth functional derivatives $F^{(n)}$ and with only a finite number of non vanishing $F^{(n)}$.
If we indicate by $F^{(n)}(0)$ the functional derivatives of $F$ we notice that $F^{(n)}(0):= F_n$.

The set $\mA(M)$, equipped with the pointwise product becomes a topological $*-$algebra with the complex conjugation taken as the $*-$operation and whose topology is induced by the one of the compactly supported smooth functions.
Such algebra is considered to be ``off shell'' because no information about the equations of motion has been employed in its construction. At the end, the equation of motion needs to be implemented singling out  the ideal $\mI$ generated by elements $P_{x_i}f(x_1,\dots, x_n)$, 
where $P_{x_i}$ is the operator realizing the equation of motion applied on the variable $x_i$, we shall call this last case ``on shell''.

The quantization of $\mA(M)$ is realized introducing a particular star product on $\mA(M)$, namely let $F$ and $G$ be two elements of $\mA(M)$
\beq\label{star}
F\star G :=  \sum_n  \frac{i^n}{2^n n!}  \langle F^{(n)},E^{\otimes n}G^{(n)} \rangle
\eeq
where $E$ is the causal propagator (the advanced minus the retarded fundamental solution).
Furthermore, also $(\mA(M),\star)$ is a topological $*$-algebra with the complex conjugation as $*$-operation.
%

%

In order to analyze the backreaction we need to deal with the stress tensor and similar fields.
Hence, we need to introduce pointwise products of fields (like the one involved in the stress tensor) in the algebra of observables. Unfortunately this is usually not possible in $\mA(M)$, because local fields, like $ \langle f(x)\de(x,y),\varphi(x)\varphi(y) \rangle$, cannot be multiplied with respect to the product rule given  in \eqref{star}.
Yet, since this kind of fields are needed in our discussion, we shall proceed as follows, first of all we isomorphically deform the algebra
and then we enlarge the deformed one by including these and other more involved fields.
The first step consists in the deformation of the $\star$ product introduced in \eqref{star} to the so-called $\star_\cH-$product. The latter is realized substituting  in \eqref{star} $E$ (the causal propagator) with $-2 i \cH$. 
The new algebra  $(\mA(M),\star_\cH)$ turns out to be isomorphic to the old one; further details about this isomorphism can be found in \cite{BF}.
After the deformation, we can safely enlarge the algebra $(\mA(M),\star_\cH)$ to $(\mF(M),\star_\cH)$ in such a way that more general local fields like the components of the stress tensor are included in $(\mF(M),\star_\cH)$.
To be more precise, $\mF(M)$ is defined as the set of functionals over smooth field configurations, i.e., they are infinitely often differentiable and every $F\in\mF(M)$ has symmetric functional derivatives $F^{(n)}(0)\in \cE'(M^n)$ (the set of distributions with compact support). Their wave front set satisfies   
$$
WF(F^{(n)}(0)) \cap  \at \overline{V}^+\cup \overline{V}^- \ct = \emptyset \;,
$$
where $\overline{V}^\pm$ are the subsets of the cotangent space $T^*(M^n)$ formed by elements whose fiber components are future, respectively past, pointing causal or null vectors. 

We proceed now to discuss the form of the states and their interplay with the previously discussed deformation.
Following the discussion presented in \cite{BDF, BF}, when we consider the enlarged algebra, we have to restrict the class of admissible states to the one satisfying the Hadamard condition.
Furthermore, the deformation must not be visible on the expectation values of the original algebra. 
Thus, in order to preserve the expectation values of the elements of $\mA(M)$ under the deformation, the states on $(\mA(M),\star_\cH)$
are constructed out of those on $(\mA(M),\star)$ by means of the push-forward induced by the isomorphism between the two algebras.  
Let us give a closer look to the effects of this push-forward on the $n-$point functions for the deformed algebra.
To this end, let us fix a state $\bom$ on $(\mA(M),\star)$, then we shall indicate by $:\om_n:$ the $n-$point function used in evaluating the elements of $(\mA(M),\star_\cH)$ and we shall call them 
{\bf regularized $n-$point functions}.
Notice that all $:\om_n:$ are obtained from the non regularized ones by the generating formula \cite{BF00,HW01}
$$
:\om_n:(x_1,\dots,x_n)\; := \left.\frac{1}{i^n} \frac{\de^n}{\de f(x_1)\dots \de f(x_n)} \om\at \exp \at \frac{1}{2}\cH(f\otimes f) -i\varphi(f)    \ct\ct \right|_{f=0}
$$
where eventually the $\bom\at\varphi(x_1)\cdot \dots \cdot\varphi(x_n)\ct$, needs to be substituted by $\om_n$.
Notice that, if $\bom$ satisfies the microlocal spectrum condition, then the all the $:\om_n:$ are smooth functions, hence the state $\bom$ can be safely extended to $(\mF(M),\star_\cH)$.
Furthermore, the regularization employed in this paper coincides with the push forward of the state under the deformation described above, in fact it holds that $:\om_2:=\om-\cH$.
%
Finally, we stress once more that, 
the stress tensor $T_{\mu\nu}$ and $\phi^2$ constructed in the main text of this paper are elements of $\mF(M).$




\subsection{Functional derivatives of $a$ and $\tau$}\label{A1}

In this subsection we shall derive some estimates, for the functionals $a$ and $\tau$,
needed  to establish the proofs of the main theorems.
Let us start by recalling the form of the scaling factor $a$ as a functional of $X$.
Fixing the initial condition $a_0$ at the cosmological time $t_0$ and its first functional derivative we can write
$$
a\aq X  \cq(t) := a_0 \exp \ag - \int_{t}^{t_0} X^{-1}(t') dt'  \cg\;,\qquad 
Da\aq X,\DX \cq(t) :=    a\aq X  \cq(t) \int_{t}^{t_0}X^{-2}(t') \DX(t')  dt'
$$
and, per direct inspection, we have the following estimate valid for every $X$ in $\cB_c$ and $t\in[0,t_0]$, 
\beq\label{Da}
\left| Da\aq X,\DX \cq(t) \right|  \leq  \left| a\aq X\cq(t)   \at\frac{1}{t}-\frac{1}{t_0}\ct   \right|  \sup_{t'\in[0,t_0]}\left| X^{-2}(t')t'^2 \right|
\| \DX \|_\infty \; \leq \frac{a_0}{t_0} \frac{1}{c^2} \|\DX \|_\infty .
\eeq
Furthermore, out of both the preceding inequalities and the definition of functional derivative as a directional derivative, it holds that  
$$
\left| a_1(t) - a_2(t) \right| = \left| \int_0^1  Da\aq {X_\la},\DX \cq(t)               
d\la
\right|
\leq
\frac{a_0}{t_0} \frac{1}{c^2} \| \DX \|_\infty, \qquad \forall t\in[0,t_0]
$$
where $\DX = {X_2}-{X_1}$ and ${X_\la}={X_1}+\la \DX$ and the last inequality descends from \eqref{Da} together with the proposition \ref{someinequalities}. 

\noindent
Let us continue analyzing the form of the conformal time as a functional of $a$ and the first functional derivative of $\tau\circ a [X]$
$$
\tau[a](t) := \tau_0-\int_t^{t_0}\frac{1}{a} dt' \;, \qquad
D\at \tau \circ a  \ct\aq X,\DX \cq =
D\tau \aq a, Da \aq X,\DX  \cq   \cq
$$
and it satisfies the following inequality 
$$
\left|
D\at \tau \circ a  \ct\aq X,\DX \cq(t) 
\right| \leq
2 \int_{t}^{t_0} \frac{1}{a} \at \frac{1}{t'} -\frac{1}{t_0}  \ct dt'  
\sup_{t'\in[0,t_0]}\left| X^{-2}(t')t'^2 \right|
\| \DX \|_\infty \;.
$$
Hence, multiplying by a sufficiently large power of $t$m we obtain that the inequalities,
\beq\label{Dtau}
\left|
t^{1/c} D\at \tau \circ a  \ct\aq X,\DX \cq (t)
\right| \leq
C(a_0,t_0,c) \;  
\sup_{t'\in[0,t_0]}\left| X^{-2}(t')t'^2 \right|\;
\| \DX \|_\infty \;  \leq  \frac{2\; t_0^{\frac{1}{c}}}{a_0\; c}\; \|\DX\|_\infty  \;,
\eeq
hold in the ball $\cB_c$ constructed out of $c$ as introduced in the definition \ref{Bc}.

\subsection{Bounds satisfied by $\chi_k$ and its functional derivatives}
\label{stimechi}

In the following we shall derive some useful estimates satisfied by the functions $\chi_k$ constructed perturbatively in \eqref{seriechi}, this  appendix is meant as a completion of the estimates presented in proposition \ref{chiconvergence}, where the convergence of \eqref{seriechi} was proven.
All the forthcoming formulas are meant to be valid for every spacetime that arises from an element $X$ of $\cB_c$ introduced in definition \ref{Bc}, along the lines of proposition \ref{functorialspacetime} (Notice that they holds also on $\cB_d'$ introduced in the proof of theorem \ref{esitenzaaltristati}).
For the same reason $t\in[0,t_0]$ everywhere. 
For our later purpose, in order to shorten some formulas,
it is useful to introduce the following two functionals
\beq\label{F1}
F_1[a](t):=\int_{0}^t a(t') \frac{m^2}{k} dt' \;, 
\qquad
F_2[a](t):=\int_{0}^t (\tau(t)-\tau(t')) a(t') m^2 dt' 
\eeq
that satisfy $|F_1[a](t)|\leq \frac{a_0}{t_0\; k} t^2 m^2 $ and $|F_2[a](t)|\leq m^2 t^2$.
Following a procedure similar to the one presented in the proof of proposition \ref{chiconvergence}, we can derive the following  estimates
\beq\label{chiN}
\left|\chi_k^{>n}[a](t)\right|
\leq 
\frac{1}{\sqrt{2k}}(F_{1/2}[a](t))^{n+1} \exp {F_{1/2}[a^2](t)}  \;.
\eeq
In the remaining part of this appendix we shall analyze the 
functional derivatives of $\chi_k-\chi_k^0$ with respect to $X \in \cB_c$; more precisely we shall compute 
$$
D(\chi_k-\chi_k^0)\left[ X,\DX\right]= D_a(\chi_k-\chi_k^0)\left[a,\tau,Da\left[ X,\DX\right]\right]
+ D_\tau(\chi_k-\chi_k^0)\left[a,\tau,D\tau\left[ X,\DX\right]\right] \;.
$$
Let us start from 
$D_a(\chi_k-\chi_k^0)[a,\tau,Da[X,\DX]]$, which, exploiting the properties of the recursive series \eqref{seriechi}, the form of $Da[X,\DX]$ and the identity \eqref{partition}, turns out to satisfy the following inequality 
$$
\left|D_a(\chi_k-\chi_k^0)\aq a,\tau,Da\at X,\DX\ct \cq(t)\right|\leq \frac{e^{F_{1/2}(t)}}{k\;\sqrt{2k}}\; m^2
\; t\; \|Da \|_\infty
$$
where $\|\cdot\|_\infty$ is the uniform norm on $C[0,t_0]$,
while $D_a\chi_k^0 = 0$.
The second contribution is more laborious; let us consider the recursive sum and let us concentrate on the contribution to the $n$-th order which looks like
\beq\label{ordinen}
\int_{0\leq t_n\leq t_{n-1} \dots \leq t}   \frac{e^{ik\tau(t_n)}}{\sqrt{2k}} \prod_{i=1}^n \frac{\sin(k(\tau(t_{i-1})-\tau(t_{i})))}{k} m^2 a(t_i)  dt_1\dots dt_n 
\eeq
where, only in the previous formula, $t_0$ corresponds to the time $t$. The action of the $\tau$-functional derivatives on \eqref{ordinen}
is twofold: On the one hand, we have to derive every factor of the form $\sin(2k(\tau-\tau'))$ and, on the other hand, we have to derive  $\tau$ in $\chi^0_k$.
Let us first of all consider the functional derivatives of $\sin(2k(\tau-\tau'))$ and, using \eqref{partition}, we obtain that the $n-$th order is bounded by 
$$
\frac{1}{(n-1)!} \frac{1}{\sqrt{2k}} \;
F_{1/2}^{n-1}(t)
\cdot \int_0^t  m^2 \frac{4}{c\;a_0}
\at\frac{t_0}{t'} \ct^{\frac{1}{c}}
 a(t')  dt' \cdot  \|\DX\|_\infty
$$
where $\|\cdot\|_\infty$ is the uniform norm on the elements of $\mB$ and 
where we have used the following estimate descending from \eqref{Dtau}
$$
\left|
D\tau \left[X,\DX\right](t_1)-D\tau\left[X,\DX\right](t_2)
\right|
\leq        \frac{4}{c\;a_0}
\at\frac{t_0}{t_2} \ct^{\frac{1}{c}}
 \|\DX\|_\infty,
\qquad  t_1 > t_2 \;.
$$
The absolute value of the functional derivative of $e^{ik\tau}$ in \eqref{ordinen} is also bounded by a similar expression
$$
\frac{1}{(n-1)!} \frac{1}{\sqrt{2k}} \;
F_{1/2}^{n-1}(t)
 \cdot \int_0^t  m^2 \frac{2}{c\;a_0}
\at\frac{t_0}{t'} \ct^{\frac{1}{c}}
 a(t')  dt' \cdot \|\DX\|_\infty\;.
$$
Both these last two estimates can be summed in order to give an exponential. Hence
$$
\left|D_\tau(\chi_k-\chi_k^0)\left[X,\DX\right](t)\right| 
\leq \frac{\exp{F_{1/2}(t)}}{\sqrt{2k}}
 \frac{6\;m^2}{2c-1} t_0  
  \|\DX\|_\infty 
$$
and, together with the $a$-functional derivative, we obtain
\beq\label{Dchimenochi0}
\left|
D(\chi_k-\chi_k^0)\left[X,\DX\right](t)
\right| 
\leq 
\aq
\frac{1}{k}\frac{a_0}{t_0} \frac{1}{c^2}
+
 \frac{6\;m^2}{2c-1}
\cq
\frac{\exp{F_{1/2}(t)}}{\sqrt{2k}} m^2  t_0   \|\DX\|_\infty \;.
\eeq
A further inequality we would like to present in this subsection is the following
\beq
\label{Dchi0Resto}
\left|(\chi_k-\chi_k^0)\left[X\right](t) \cdot D\overline{\chi_k^0}\left[X,\DX\right](t) \right|
\leq
\frac{\exp{F_{1/2}(t)}}{{2k}}
 \frac{2\;m^2}{2c-1} t_0  
  \|\DX\|_\infty 
\eeq
which can be shown to hold in a similar way as the bound found above for the functional derivative of $e^{ik\tau}$.
We conclude this subsection with two inequalities, valid for $n\geq 0$, that can be shown to hold in a similar way as before:
\beq\label{Dchin}
\left|D\chi_k^{>n}\left[X,\DX\right](t)\right| \leq F_{1/2}^{n}(t)
 \frac{\exp{F_{1/2}(t)}}{\sqrt{2k}}
 \aq
\frac{1}{k}\frac{a_0}{t_0} \frac{1}{c^2}
+
 \frac{6\;m^2}{2c-1}
\cq
m^2t_0
  \|\DX\|_\infty 
\eeq	
and
\beq\label{chinDchi0}
\left|\overline{\chi_k^{>n}}\left[X\right](t) \cdot D\chi_k^{0}\left[X,\DX\right](t) \right| \leq
F_{1/2}^n(t)\frac{\exp{F_{1/2}(t)}}{{2k}}
 \frac{2\;m^2}{2c-1} t_0  
  \|\DX\|_\infty \;.
\eeq

\subsection{Proof of Proposition \ref{a2phi2bound}}\label{stimephi}
\begin{proof}
The proof is done dividing the expression for $\Phi$ in a finite number of parts and showing that every contribution separately satisfies the desired inequalities with respect to some constants that have the properties stated in proposition \ref{a2phi2bound}.
In the next, since most of the computations are rather long but straightforward, we shall concentrate only on the key points, and we shall make extensive use of all the estimates derived in the previous appendix.
Throughout this proof we shall make use of proposition \ref{functorialspacetime} in order to associate a spacetime to an element $X$ of $\cB_c$. According to the same proposition, we shall 
indicate $X^{-1}$ as $H$, and the $t$ which appears in some estimates  has to be thought as being contained in the interval $[0,t_0]$.
Let us start dividing the integral over $k$ in two parts 
$$
\Phi \left[X\right] = \int_0^1 dk k^2   \aq \overline{\chi_k} \chi_k -\overline{\chi_k^0} \chi_k^0 \cq   +     \int_1^\infty dk k^2  \aq \overline{\chi_k} \chi_k -\overline{\chi_k^0} \chi_k^0  +\frac{a^2 m^2}{4k^3}\cq  - m^2a^2 \log(am)\;,
$$
where the regularization is needed only for large $k$, whereas $\chi$, $a$ and $\tau$ are functionals of $X$ as discussed previously.
Notice that, the zeroth order in $m^2$, present in $\langle\phi^2\rangle$ through $\overline\chi_k^0\chi_k^0$, vanish due both to the discussion about the regularization freedom and to the requirement that Minkowksi spacetime, with respect to the Minkwoksi vacuum, is a solution of the problem.
For this reason we have subtracted it in \eqref{Phi} and we shall not consider it anymore.
%
%
Let us start considering the first part of the integral, to which we shall refer to as the infrared part and we shall indicate it by $\Phi^I$.
In this respect we shall write
$$
\aq \overline{\chi_k} \chi_k -\overline{\chi_k^0} \chi_k^0 \cq 
=
\at\overline{\chi_k}-\overline{\chi_k^0}\ct\at\chi_k-\chi_k^0\ct
+
\at\overline{\chi_k}-\overline{\chi_k^0}\ct\chi_k^0
+
\overline{\chi_k^0}\at\chi_k-\chi_k^0\ct
$$
and, hence, using the estimates \eqref{chiN} for 
$\chi_k$ and the one for the functional derivatives 
\eqref{Dchimenochi0} and \eqref{Dchi0Resto}, with $F_2$ as in \eqref{F1}, 
we obtain 
$$
\left|\Phi^I(t)\right| \leq \frac{1}{4}\aq  F_2(t)^2 \exp {(2\cdot F_2(t))} 
+2F_2(t) \exp{F_2(t)}\cq  \leq \frac{1}{4} \at m^4t^4 + 2 m^2 t^2   \ct e^{2m^2t^2}
$$
where $F_2(t) \leq m^2t^2$, as it can be seen by direct inspection of the definition of both $F_2(t)$ and $\tau$, and
$$
\left|D\Phi^I\left[X,\DX\right](t)\right| \leq  
\ag
\aq F_2(t) \exp {F_2(t)} +1 \cq\aq
\frac{a_0}{t_0} \frac{1}{c^2}
+
 \frac{3\;m^2}{2c-1}
\cq
+
\frac{1\;m^2}{2c-1}
\cg 
\exp{F_{2}(t)}\; m^2  t_0   \|\DX\|_\infty \;.
$$
This conclude the analysis of the contribution of the lower energies to the value of $\Phi$, we can now proceed to analyze the higher frequencies which we shall indicate as $\Phi^U$.
We shall split it in powers of $m^2$ and discuss the first three powers separately. Afterwards, before summing everything together, we shall analyze the remainder.

\vsp
\noindent
{\bf I) order.} Let us start considering the first order in $m^2$ and its contribution to $\Phi^U$ in particular, 
$$
\Phi^U_1[a,\tau](t) := 
\int_{1}^\infty k^2
\aq \overline {\chi_k^1} \chi_k^0[a,\tau](t) 
+\overline {\chi_k^0} \chi_k^1[a,\tau](t) 
+
\frac{a^2(t) m^2}{4k^3}\cq dk\;  - m^2a^2(t) \log(a(t)m)\;
$$
which can be expanded as 
\begin{gather*}
\Phi^U_1[a,\tau](t) := 
\int_{1}^\infty dk\;
 \frac{m^2}{4 k}
 \int_{0}^t 
   \cos{(2 k(\tau-\tau'))} 
  \frac{\pa a^2}{\pa t}(t')\;dt'  
 - m^2 a(t)^2
\log \at m\; a(t)\ct  
\end{gather*}
where both $a$ and $\tau$ have to be thought as functionals of $X$
 and where we have written the perturbative  series \eqref{seriechi} using $t$, instead of $\tau$, as the time variable. Furthermore, $\tau$ has to be intended as $\tau[X](t)$ while $\tau'$ as $\tau[X](t')$.
Let us rewrite it in the following form 
 \begin{gather*}
\Phi^U_1[a,\tau](t) := 
\int_{1}^\infty dk\;
 \frac{m^2}{8 k^2}
 \int_{0}^t 
   \sin{(2 k(\tau-\tau'))} 
  \frac{\pa}{\pa t} \at a \frac{\pa a^2}{\pa t}   \ct(t')  \;dt'  
 - m^2 a(t)^2
\log \at m a(t)\ct  
\end{gather*}
and, 
using the fact that the $t$ derivative of $X$ is positive, 
we can obtain the following bound 
$$
|\Phi^U_1(t)| \leq  m^2 a^3(t) X^{-1}(t) + m^2 a(t) t +m^2 a^2(t) \log(ma(t))\;.
$$
Let us proceed to analyze the functional derivative
$$
D\Phi^U_1[a,\tau,\DX]= D_a\Phi^U_0[a,\tau,Da[X,\DX]] +
D_\tau\Phi^U_0[a,\tau,D\tau[X,\DX]], 
$$ 
and let us analyze both term separately, first of all,
setting $F:=Da[X,\DX]$,
\begin{gather*}
D_a\Phi^U_1[a,\tau,F](t)  := 
\int_{1}^\infty 
 \frac{m^2}{8 k^2}
 \int_{0}^t 
   \sin{(2 k(\tau-\tau'))} 
  \aq
  F(8a^2H^2+4a^2\dot{H})+
  8\dot{F} a^2 H+ 2 a^2 \ddot{F} 
  \cq(t')
  \;dt'\;dk -\\
   - 2 m^2 a(t) F
\log \at m a(t)\ct
- m^2 a(t) F \;,
\end{gather*}
where as usual $H=X^{-1}$ and the dots indicates the derivatives with respect to the cosmological time $t$.
Using the estimates present in proposition \ref{someinequalities} and \eqref{Da}, it is now an easy task to obtain the desired bounds.
Notice in particular that the second order time derivative of $F$ multiplied by $a^2$ can be estimated by $\|\DX\|_\mB$ (the norm of $\mB$), as it can be verified directly from the definition of $a[X]$. 
Let us proceed with the second functional derivative, and now setting 
$G:=D\tau[X,\DX]$ we obtain
 \begin{gather*}
D_\tau\Phi^U_1[a,\tau,G](t)  := 
\int_{1}^\infty \;
 \frac{m^2}{4 k}
 \int_{0}^t 
   \cos{(2 k(\tau-\tau'))}\; \aq G(t)-G(t')  \cq \;
\aq 6a^3H^2+2a^3\dot{H}\cq(t')    
  \;dt' \;dk \;.
\end{gather*}
We could in principle integrate by parts in order to obtain an extra $1/(2k)$ factor. 
Unfortunately this would not help much, since, due to the extra $t$-derivative the left hand side would depend on $\ddot{H}$ and we do not have control on the second $t$-derivative of  $H$ in $\cB_c$. 
At the same time we have to remember that $\frac{\cos{k x}}{k}$ is integrable in $k$ on $[1,\infty)$ for every $x>0$ and that the $k$-integral present above needs to be understood in the distributional sense, hence taken with the following $\epsilon$-prescription 
 \begin{gather*}
D_\tau\Phi^U_1[a,\tau,G](t')  := 
\lim_{\epsilon=0^+}
\int_{1}^\infty \!\!\! e^{-\epsilon k}
 \frac{m^2}{4 k}
 \int_{0}^t 
   \cos{(2 k(\tau-\tau'))} \aq G(t)-G(t')  \cq 
(6a^3H^2+2a^3\dot{H})(t')    
  \;dt'\;dk\;.
\end{gather*}
Since 
$\aq G(t)-G(t')  \cq \;
(6a^3H^2+2a^3\dot{H})$ is $L^1(0,t)$,
at fixed $\epsilon$, the absolute value of the integrand in $dk\wedge dt$ is integrable. We can thus switch the order of integration to obtain
 \begin{gather}
D_\tau\Phi^U_1[a,\tau,G](t)  := 
m^2 \int_{0}^t 
\aq G(t)-G(t')  \cq \;
(6a^3H^2+2a^3\dot{H})(t')\; 
 \lim_{\epsilon=0^+}
\int_{1}^\infty\!\!\! 
 \frac{e^{-\epsilon k}}{4 k}
   \cos{(2 k(\tau-\tau'))}\;     dk
  \;dt'\;.
  \label{cosxx}
\end{gather}
We are now ready to perform the integral in $k$ and to notice that the result is a smooth function everywhere in  $\tau-\tau'$ but in $0$ where a logarithmic divergence in $\tau-\tau'$ appears when the limit $\epsilon=0^+$ is computed. Let us call $L(\tau-\tau')$ the result of the $k$-integration in the limit $\epsilon=0$.
Notice that $L(\tau(t)-\tau'(t'))/a(t')$ is absolutely integrable for $t'$ in $[t-\de,t]$ and, moreover, we can adjust $\delta$ in such a way both that this integral is equal to $1$ and that 
 $|L(\tau(t)-\tau'(t'))|\leq 1 $  for $t'$ in  $(0, t-\delta)$.
Hence, 
since $a(t') \aq G(t)-G(t')  \cq \;
(6a^3H^2+2a^3\dot{H})(t')$ is uniformly bounded for $t'$ in $[0,t]$, we can split the $t'$-integral in  \eqref{cosxx} in two parts, namely over $[0,t-\delta ]$ and over $[t-\delta,t]$. Both can be uniformly estimated and finally the following absolute bound for $\left| D_\tau\Phi^U_1\right|$ can be established 
$$
\left| D_\tau\Phi^U_1(t)\right| \leq \sup_{t'\in[0,t]}\left| a(t') \aq G(t)-G(t')  \cq \right|\;
\| 6a^3H^2+2a^3\dot{H}  \|_\infty + \int_0^t \left|
\aq G(t)-G(t')  \cq \;
(6a^3H^2+2a^3\dot{H})
 \right| dt' \;.  
$$
Combining all the estimates, we obtain the desired behavior, namely that the absolute value of the functional derivatives is controlled by $ C t_0  \|\DX\|$.

\vsp
\noindent
{\bf II) order.}
The computation at the second order is more involved due to a pair of integrals which needs to be addressed, but at the same time there are no subtleties with the extra derivatives which could necessitate  control on the derivatives of the variation higher than the first one.
We start reminding the form of this term
\begin{gather*}
\Phi^U_2[a,\tau] := 
\int_{1}^\infty\; k^2 \aq
\overline {\chi_k^2} \chi_k^0[a,\tau] 
+\overline {\chi_k^1} \chi_k^1[a,\tau] 
+\overline {\chi_k^0} \chi_k^2 [a,\tau]
\cq dk\;,
\end{gather*}
which can be more explicitly rewritten as
\begin{gather}
\Phi^U_2[a,\tau](t) := 
\int_{1}^\infty \!\!\! dk\; \frac{m^4}{k}
\aq 
\aq \frac{1}{2k} \int_0^t \at \sin(k(\tau-\tau')) \ct^2 \frac{\pa a^2}{\pa t}(t') dt'  \cq^2
+ 
\frac{1}{8k}\aq\int_0^t \sin(2k(\tau-\tau')) \frac{\pa a^2}{\pa t}(t') dt'  \cq^2
\cq 
-
\nonumber
\\
-
\int_{1}^\infty dk\; \frac{m^4}{4 k^2}
\int_0^t\sin(2k(\tau-\tau')) \frac{\pa a^2}{\pa t}(t')  \int_{t'}^t a(t'') dt'' dt'
+
\int_{1}^\infty dk\; \frac{m^4}{4 k^2}
\int_0^t\sin(2k(\tau-\tau'))  a^3(t')   dt'
\label{Phi2U} \;.
\end{gather}
Out of such expression, an estimate satisfied by $|\Phi^U_2|$ can be easily derived and it looks like
$$
\left|\Phi^U_2[a,\tau](t)\right| \leq
m^4 
\aq
\frac{5}{4}(a^2H t)^2
+
\frac{1}{2}a^3 H t^2
+\frac{1}{4} a^3 t
\cq\;.
$$
Let us proceed to the analysis of the functional derivatives of $\Phi$ and let us notice that the analysis of $D_a\Phi_2$ can also be easily done yielding
$$
\left|D_a\Phi^U_2[a,\tau,Da](t)\right| \leq    
\frac{m^4}{8} \aq 6 a^3 H^2 t^2 + 3 a^2 H t^2 +t a^3  \cq \| Da \|_\infty  + 
\frac{m^4}{8} \aq 2 a^2 H t^2 + at^2 \cq \left\| a  (\dot{Da})  \right\|_\infty\;.
$$
A little bit more laborious is the handling of the $\tau$-directional derivative, because, operating as above, an extra factor $2k$  appears in the formula. Furthermore, it could make the $k$-integral divergent when the trigonometric functions are replaced with one in the maximization procedure of the second, third and fourth factor in \eqref{Phi2U}.
In order to avoid this problem, we can integrate by parts after performing the functional derivative, along the following lines:   
\begin{gather*}
D_\tau \int_0^t \sin(2k(\tau-\tau'))  \frac{\pa a^2}{\pa t}  dt' =
\int_0^t  \sin(2k(\tau-\tau'))  \frac{\pa}{\pa t}\aq a \at\frac{\pa a^2}{\pa t}\ct D(\tau-\tau')    \cq dt'.
\end{gather*}
The price to be paid is in the extra time derivative, we have to deal with. In fact, proceeding in that way we obtain the following estimate 
$$
\left|D_\tau\Phi^U_2[a,\tau,D\tau](t)\right| \leq 
C_1 \int_0^{t} a |D(\tau-\tau')| dt' +
C_2 \int_0^{t} a^2 \left|\frac{d}{dt'} D(\tau-\tau')\right| dt'  \;.
$$
It is now a straightforward task to obtain the desired inequalities.

\vsp
\noindent
{\bf Rest$_2$)}
The contribution of the order larger than 2 in $m^2$ is
\begin{gather*}
\Phi^U_{>2}[a,\tau] := 
\int_{1}^\infty dk\; k^2
\aq \overline{\chi^1_k} \chi^2_k +{\chi^1_k}  \overline{\chi^2_k}
+
\at \overline{\chi^0_k}+\overline{\chi^1_k}\ct \chi^{>2}_k
+ 
\overline{ \chi^{>2}_k} \at {\chi^0_k}+{\chi^1_k}\ct 
+
\overline{\chi^{>1}_k}\chi^{>1}_k 
\cq \;.
\end{gather*}
Hence, employing the inequality \eqref{chiN}, and the definition of $F_1$ in \eqref{F1}, we obtain the bounds
\begin{gather*}
\left|
\Phi^U_{>2}[a,\tau](t) 
\right| 
\leq  \int_{1}^\infty dk 
\frac{1}{k^2}    
\at {m^2} \frac{a_0}{t_0} t^2  \ct^3
\aq 
1
+
\at 
1+ \frac{m^2}{k} \frac{a_0}{t_0} t^2  
\ct
\exp{(F_1[a])}    
+
\frac{m^2}{k} \frac{a_0}{t_0} t^2
\exp{(2F_1[a])}
\cq 
\end{gather*}
where $F_1(t)\leq \frac{m^2}{k} \frac{a_0}{t_0} t^2$, more explicitly we can rewrite it as 
\begin{gather*}
\left|
\Phi^U_{>2}[a,\tau](t)
\right| 
\leq  
3\at {m^2} \frac{a_0}{t_0} t^2  \ct^3
\at 1+m^2 {a_0}{t_0} \ct
\exp{\at 2 m^2 {a_0}{t_0} \ct} \;.
\end{gather*}
We do not analyze the functional derivative of this contribution, because, due to the extra $k$ appearing in the functional derivative of the trigonometric functions with respect to $\tau$, we have to further split it.

\vsp
\noindent
{\bf III) order}
The contribution at third order in $m^2$ is
\begin{gather*}
\Phi^U_3[a,\tau] := 
\int_{1}^\infty dk\; k^2
\aq
\overline {\chi_k^3} \chi_k^0[a,\tau] 
+\overline {\chi_k^2} \chi_k^1[a,\tau] 
+\overline {\chi_k^1} \chi_k^2[a,\tau] 
+\overline {\chi_k^0} \chi_k^3 [a,\tau]
\cq
\end{gather*}
and its explicit form is 
\begin{gather}
\Phi^U_3[a,\tau](t) := 
\int_1^\infty dk  \frac{m^6}{k^2} 
\cdot 
\nonumber
\\
\aq 
\int_{0<t_3<t_2<t_1<t} \!\!\!\!\!\!\!\!\!\!\!\! 
\sin(k(\tau-\tau_1))\sin(k(\tau_1-\tau_2))\sin(k(\tau_2-\tau_3))\cos(k(\tau-\tau_3)) 
a(t_1) 
a(t_2)
a(t_3)
\;
dt_1
dt_2
dt_3
+
\right.
\nonumber
\\
+
\left.
\int_0^t \int_{0<t_3<t_2<t} \!\!\!\!\!\!\!\!\!\!\!\!
\sin(k(\tau-\tau_1))\sin(k(\tau-\tau_2))\sin(k(\tau_2-\tau_3))\cos(k(\tau_1-\tau_3)) 
a(t_1) 
a(t_2)
a(t_3)
\;
dt_1
dt_2
dt_3
\cq 
\label{terzordineexp}
\end{gather}
An estimate of the form 
$$
\left|D\Phi^U_3(t)\right| \leq  C(t_0,a_0,c)\|\DX\|_\infty
$$
holds for the first functional derivative.
Notice that, in the derivation of the preceding estimate starting from \eqref{terzordineexp},
a problem could in principle arise when the 
$\tau$ functional directional derivative is computed because in such a case an extra $k$ factor appears and the $k$ integral is logarithmic divergent.
Hence, in order to obtain a finite $k$-integration, before performing the functional derivative, we have to rewrite the product of the trigonometric functions as a sum of products of trigonometric functions depending only on $\tau-\tau_1$, $\tau_1-\tau_2$, $\tau_2-\tau_3$ for the first line in \eqref{terzordineexp}
and  on $\tau-\tau_1$, $\tau-\tau_2$, $\tau_2-\tau_3$ for the second one. Subsequently we can safely perform a couple of integration by parts.
The outcome is that the result of the $t$-integrations is of order $1/(2k)$ in $k$ and this extra $k$ permits to obtain finite bounds for the $k$-integration also for the $\tau$-functional derivative. 
The price to be paid is that, afterwards, in the $t_i$ integrations  the derivatives of $a(t)$ might appear.
This also holds for the $a$ directional functional 
 derivative which has to be carefully computed.
More precisely we obtain
$$
\left| 
D_\tau \Phi^U_3(t) \right| \leq \aq \int_1^\infty dk  \frac{m^6}{2 k^2}\cq 
 \aq 112 t^2 a^4 H + 6t a^3     \cq    \int_0^t  a| D\tau | 
$$
and
$$
\left| 
D_a \Phi^U_3(t) \right|  \leq    
\aq\int_1^\infty dk  \frac{m^6}{2 k^3}
\cq 
\aq 84 Ha^3 t^3 + 6a^3 t^2    \cq  \left\| Da \right\|_\infty 
+
\aq
\int_1^\infty dk  \frac{m^6}{2k^3} 
\cq
\aq 28 a^2 t^3 \cq  \left\| a (\dot{Da}) \right\|_\infty\;, 
$$
out of which it is straightforward to obtain the wanted inequality.

\vsp
\noindent
{\bf Rest$_3$)}
The contribution of the order larger than 3 in $m^2$ is
\begin{gather*}
\Phi^U_{>3}[a,\tau] := 
\int_{1}^\infty dk\; k^2
\aq 
 \overline{\chi^0_k} \chi^{>3}_k
 +
 \overline{ \chi^{>3}_k}{\chi^0_k}
+
\overline{\chi^1_k} \chi^{>2}_k
+
\overline{ \chi^{>2}_k}{\chi^1_k}
+
\overline{\chi^{>1}_k}\chi^{>1}_k 
\cq
\end{gather*}
and, using \eqref{chinDchi0} and \eqref{Dchin} together with \eqref{chiN}, we can find a constant $C'(a_0,t_0,c)$, with the desired properties, i.e.,
$$
\left|D\Phi^U_{>3}(t)\right| \leq C'(a_0,t_0,c)\|\DX\|_\infty ,\qquad \forall t\in[0,t_0] .
$$

\vsp
\noindent
After a rather long but straightforward computation, 
we are ready to collect all the estimates presented above hence obtaining that   
$$
\left|\Phi(t)\right|   \leq     C(a_0,c,t_0) t \;, \qquad
\left|D\Phi\aq X,\DX\cq(t)\right|   \leq     C(a_0,c,t_0)  \|\DX\|_\infty,\qquad \forall t\in[0,t_0]\;,   
$$
where $C(t_0\ell,c,t_0)$ remains bounded for small $t_0$.
The thesis of the proposition can be obtained noticing that 
$\|\DX\|_\infty \leq t_0 \|\DX\|_\mB$ in particular.
\end{proof}

\vspace{0.5cm}

\end{document}